\documentclass[10pt,aps,twocolumn,prl,groupedaddress,amsmath]{revtex4-2}
\usepackage{graphicx}
\usepackage{braket}
\usepackage[pagewise]{lineno}
\usepackage[breaklinks=true,colorlinks,citecolor=blue,linkcolor=blue,urlcolor=blue]{hyperref}
\usepackage{tabularx}
\usepackage{xcolor}

\usepackage[normalem]{ulem}

\begin{document}
\title{Entanglement-Enhanced Nanoscale Single-Spin Sensing}

\author{Xu Zhou$^{1}$$^{\S}$}
\author{Mengqi Wang$^{1}$$^{\S}$}
\author{Xiangyu Ye$^{1}$}
\author{Haoyu Sun$^{1}$}
\author{Yuhang Guo$^{1}$}
\author{Han Shuo$^{1}$}
\author{Zihua Chai$^{1}$}
\author{Wentao Ji$^{1}$}
\author{Kangwei Xia$^{1,2}$}
\author{Fazhan Shi$^{1,2}$}
\author{Ya Wang$^{1,2}$}
\author{Jiangfeng Du$^{1,2,3}$}

\affiliation{$^1$ Laboratory of Spin Magnetic Resonance, School of Physical Sciences, Anhui Province Key Laboratory of Scientific Instrument Development and Application, University of Science and Technology of China, Hefei 230026, China}
\affiliation{$^2$ Hefei National Laboratory, University of Science and Technology of China, Hefei 230088, China}
\affiliation{$^3$ Institute of Quantum Sensing and School of Physics, Zhejiang University, Hangzhou 310027, China}
\affiliation{$^{\S}$ These authors contributed equally to this work}

\date{\today}

\begin{abstract}

Detecting individual spins--including stable and metastable states--represents a fundamental challenge in quantum sensing with broad applications across condensed matter physics, quantum chemistry, and single-molecule magnetic resonance imaging. While nitrogen-vacancy (NV) centers in diamond have emerged as powerful nanoscale sensors, their performance for single-spin detection remains constrained by substantial environmental noise and restricted sensing volume. Here, we propose and demonstrate an entanglement-enhanced sensing protocol that overcomes these limitations through the strategic use of entangled NV pairs. Our approach achieves a 3.4-fold enhancement in sensitivity and a 1.6-fold reduction in spatial resolution relative to single NV centers under ambient conditions. The protocol employs carefully engineered entangled states that amplify target spin signals through quantum interference while suppressing environmental noise.
Crucially, we extend these capabilities to resolve metastable single-spin dynamics, directly observing stochastic transitions between different spin states by identifying state-dependent coupling strengths. 
This dual functionality enables simultaneous detection of static and dynamic spin species for studying complex quantum systems.
The achieved performance establishes entanglement-enhanced sensing as a viable pathway toward atomic-scale characterization of quantum materials and interface.

\end{abstract}


\maketitle 

Quantum sensing, enabled by individual nanoscale quantum sensors, is becoming a transformative technology, offering unparalleled precision in characterizing microscopic structures \cite{RMP_Du_2024,Degen_resolution_2017,Jelezko_submillihertz_2017,Joerg_Chemical_2017,Casola2018,Tim_Nspin_cluster_2019,Xu_moire_2021,Van_Spinwave_2023,Ya_Correlated_2024} and unraveling fundamental microscopic mechanisms \cite{Many-body_NRP_2024,Lukin_Graphene_2019,Norman_Hydrodynamics_2021,Lukin_Johnsonnoise_2015,Walsworth_Diracfluid_2020}. A pivotal strategy for advancing this technology lies in leveraging quantum entanglement to surpass the limitations of individual quantum sensors. While entangled quantum sensing has been extensively explored to approach the fundamental precision limits dictated by quantum physics \cite{Jones2009,Mark_100times_2016,Xie2021,Thomas_Metrology_2022}, the practical integration of quantum entanglement into nanoscale quantum sensing systems with superiority remains a significant and ongoing challenge. This challenge primarily stems from the inherent fragility of quantum entanglement, particularly for quantum sensors operating in close proximity to the interfaces or material surfaces, which results in severe decoherence\cite{Jelezko_surface_2014,Degen_surfacenoise_2014,Jayich_surfacenoise_2014,Barry2020}. The resulting ultra-short entanglement coherence time can lead to even worse performance than single-nanoscale quantum sensors, especially when accounting for the overhead associated with entangled state preparation and detection \cite{Cooper2019}.

\begin{figure*} [ht]
	\begin{center}
	\includegraphics[scale=0.95]{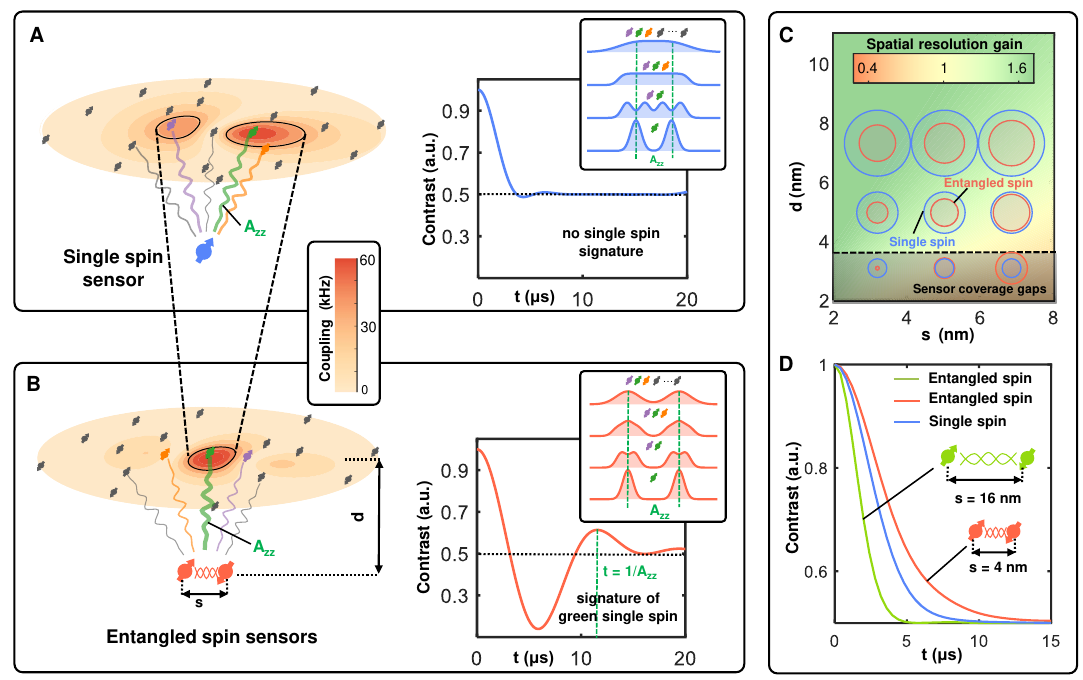}
	\end{center}
	\caption{ Schematic of entangled nanoscale sensing with spin sensors. $\textbf{A.}$ Nanoscale single-spin sensing within a substantial noise from the interface spin bath (Left). The red regions highlight optimal locations for single-spin target detection, where the coupling strength $\mathrm{A_{zz}}$ is dominant, thereby determining the spatial resolution (Methods). Spectral crowding, caused by coupling to multiple spin targets and the limited coherence time of the spin sensor, prevents reliable discrimination of single-spin signals (Right). $\textbf{B.}$ Entangled nanoscale sensing with two strongly coupled spin sensors seperated by a horizontal distance of s (Left). As the spatial resolution (red regions) improves, and the coupling strength selectively amplifies, a single-spin oscillation signature is successfully observed in the coherence evolution of the entangled sensors (Right). $\textbf{C.}$ Spatial resolution dynamics for sensor configurations in different depths and separations. The size of the blue circles (single spin sensor) and red circles (entangled sensors) indicates the corresponding spatial resolution. The red (blue) region marks entangled nanoscale sensing with increased (reduced) spatial resolution relative to the single spin sensor. $\textbf{D.}$ Entangled states $\frac{1}{\sqrt{2}}(\ket{\uparrow \downarrow}+\ket{\downarrow \uparrow})$ show distinct free-induction time for close and large sensor seperations. The results are calculated using $\mathrm{d} = 9\ \mathrm{nm}$, $\mathrm{s} = 4\ \mathrm{nm}$ (red line), $\mathrm{s} = 16\ \mathrm{nm}$ (green line), and interface spin density of 0.01 $\mathrm{nm}^{-2}$ by averaging over 1000 randomly generated surface spin distributions.
 }
	\label{Fig1}
\end{figure*}

Here we present a general approach to suppressing environmental noises while maintaining sensitivity to the target single spin. The key idea is to develop strongly coupled quantum sensors positioned in close proximity to each other, enabling superior entangled nanoscale quantum sensing (Fig.\ref{Fig1}A and B). 
By preparing entangled states in subspaces with an average zero quantum number (such as $\frac{1}{\sqrt{2}}(\ket{\uparrow \downarrow}+\ket{\downarrow \uparrow})$, $\uparrow(\downarrow)$ denotes $m_s=+\frac{1}{2}(-\frac{1}{2})$ for spin-1/2 system), the sensors exhibit reduced coupling strength across most of the surface (blue region in Fig.\ref{Fig1}B), making them effectively immune to common surface noise. Importantly, the sensing system also retains the sensitivity to single-spin targets located in confined surface regions (red region in Fig.\ref{Fig1}B), achieving enhanced spatial resolution. 

\begin{figure*} [ht]
	\begin{center}
	\includegraphics[scale=0.94]{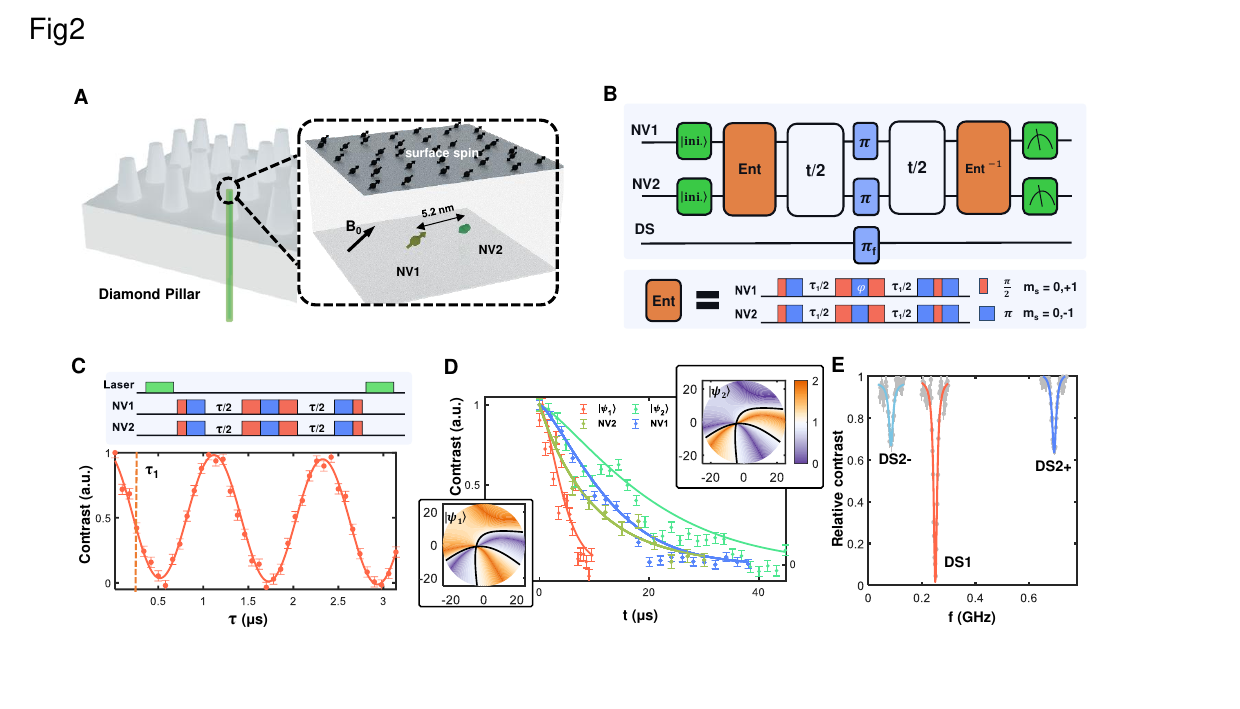}
	\end{center}
	\caption{Entangled-states-based spectroscopy of individual dark spins. $\textbf{A.}$ Schematic diagram of nanopillar structures with NV pairs in its central region. $\textbf{B.}$ DEER spectroscopy of dark spin (DS) targets utilizing NV1-NV2 entangled states. The selective coupling of DS targets to NV1-NV2 entangled states is achieved by a functional $\pi$ pulse (DS) with the corresponding resonant frequency $f$. Ent denotes the standard pulse sequence for preparing entangled states. $\varphi = 0$ for $\ket{\psi_1}$, and $\varphi = \pi/2$ for $\ket{\psi_2}$. $\textbf{C.}$ Ent sequence with varying evolution time $\tau$ determines both the coupling strength $A_{\mathrm{NV1,NV2}} = 414(3)\mathrm{kHz}$ between NV1 and NV2 and the optimal evolution time $\tau_1=1/8A_{\mathrm{NV1,NV2}}$ for preparing entangled states. $\textbf{D.}$ Spin coherence decay for both separated states (NV1 and NV2) and entangled states ($\ket{\psi_1}$ and $\ket{\psi_2}$) measured using Hahn echo sequences. The calculated coupling strength of $\ket{\psi_1}$ and $\ket{\psi_2}$ to the surface spin bath shows distinct spatial patterns, with enhanced (red) and suppressed (blue) regions relative to single-spin coupling, directly explaining the state-dependent coherence times. 
    $\textbf{E.}$ DS frequency scanning using $\ket{\psi_2}$ revealed the distinct spectral peaks of dark spins under $\mathrm{B_0}=89$ Gauss.} 
	\label{Fig2}
\end{figure*} 

A critical requirement for this entangled sensing scheme, preserving all the advantages of entanglement, is precise control over the locations of the quantum sensors to ensure their separation ($\mathrm{s}$) is small relative to their depth ($\mathrm{d}$) from the interface bath. Fig.\ref{Fig1}C and D illustrate the performance in spatial resolution and spin coherence time at different separations. At large separation, each quantum sensor interacts with an independent spin bath, leading to a degraded spatial resolution and enhanced docoherence rate for the entangled state. In contrast, at relatively smaller separations, a 1.6-fold improvement in spatial resolution is achieved (Fig.\ref{Fig1}C). Combined with the enhanced spin coherence time, entangled nanoscale sensing demonstrates superior performance in detecting individual spin targets' coherent monotonic oscillation signature within a dense spin bath (Fig.\ref{Fig1}B Right). This scheme can be readily implemented in various nanoscale sensors, such as spin defects in silicon-carbide \cite{Awschalom_SiC_2011,Joerg_SiC_2024} and single molecule spin probes\cite{Benjiamin_molecule_2024}, offering superior performance across a wide range of applications. 

We demonstrate the scheme using nitrogen-vacancy (NV) centers in diamond, a leading nanoscale quantum sensor widely applied in biology, condensed matter, and material science \cite{Balasubramanian2008,Maze2008,Renbao_nspin_2012,Staudacher2013,Rugar2014,Shi2015,Sushkov2014,Kucsko2013, Degen2017,Ya_Highpressure_2024}. An NV pair oriented along different crystallographic axes is fabricated within the optimal performance locations of nanopillars that can function as a scanning magnetometry tip \cite{Du_AFM_2021} (Fig.\ref{Fig2}A). 
The fabrication process utilizes an isotopically pure diamond consisting of 99.999\% $^{12}$C \cite{Han2025} and employs nanometer precision self-aligned techniques for nitrogen ion implantation \cite{Wang2022}. This yields strongly coupled NV centers with an average depth of 20(7) nm for nanoscale sensing. This shallow depth results in a spin coherence time $T_{2}$ of 14.5 and 10.2 $\mu$s for NV1 and NV2, respectively, which are dominated by the diamond surface noise \cite{Han2025}.
Under a static magnetic field, $\mathrm{B_0} = 89$ G aligned along the NV1 axis, double electron-electron resonance (DEER) measurements reveal a strong coupling strength of 414(3) kHz, corresponding to a 5.2(3) nm separation between two NV centers (Fig.\ref {Fig2}A, Extended Data. 

\begin{figure*} [ht]
	\begin{center}
	\includegraphics[scale=0.85]{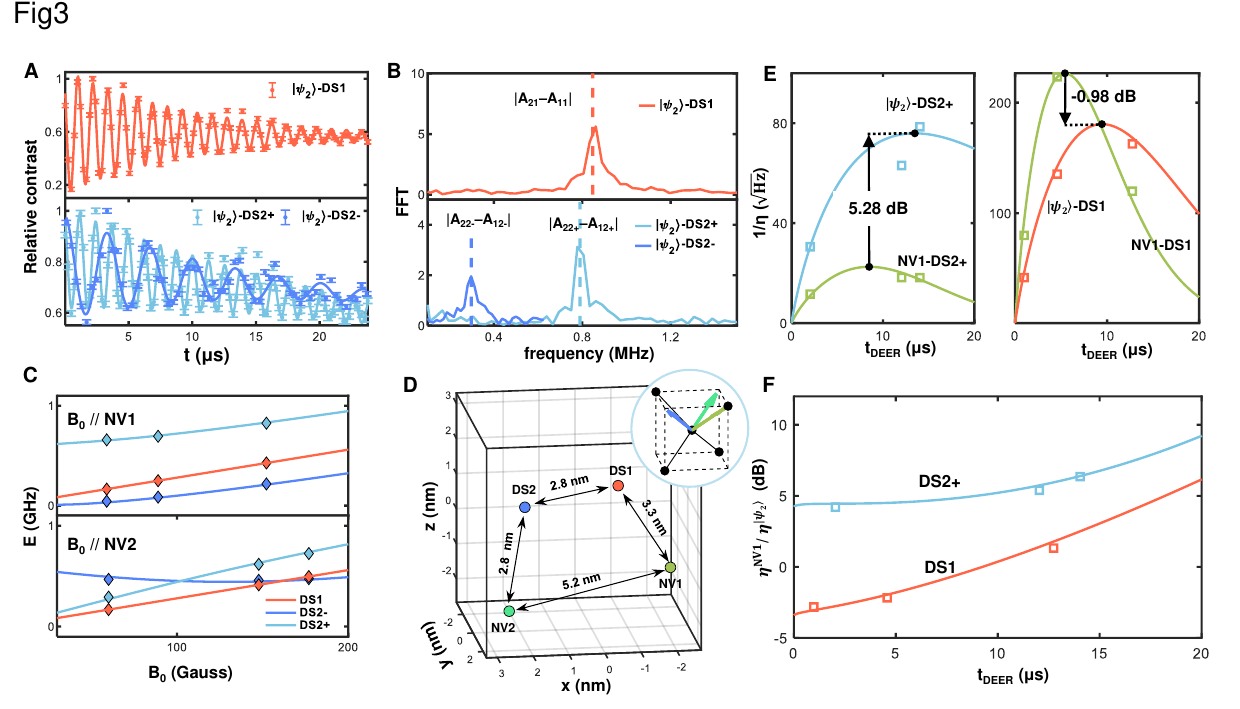}
	\end{center}
	\caption{Stable dark spin imaging with enhanced detection sensitivity via quantum entanglement. $\textbf{A.}$ Coherent phase evolution in $\ket{\psi_2}$ by coupling to individual dark spins. The oscillation frequency depends on the specific coupling strength. A shorter coherence time (Top) than detecting DS2 (Bottom) is due to the simultaneous inversion of the spin-1/2 bath on the diamond surface in detecting DS1. $\textbf{B.}$ The corresponding Fast Fourier Transform (FFT) for signals in $\textbf{A}$. The dotted line represents the calculated coupling difference between two NV centers and target spins according to the spatial mapping results in $\textbf{D}$. The $\ket{\psi_2}$ enhances the coupling strength of DS2 while diminishing that of DS1. $\textbf{C.}$ Identification of spin number and Hamiltonian via the $\ket{\psi_2}$-based spectroscopy of dark spins under different external magnetic fields. The solid lines are the calculated resonance frequencies for spin-1/2 (red) and spin-1 (blue and light blue) (see SI). $\textbf{D.}$ Dark spins imaging utilizing the dipole-dipole interaction dependence on external magnetic fields (Extended Data) as in $\textbf{B}$ and the spin property in $\textbf{C}$. The upper right subplot displays the determined principal axis orientations of NV1, NV2, and DS2. $\textbf{E.}$ Sensitivity analysis for detecting DS1 and DS2 via $\ket{\psi_2}$. The left and right plots depict the reciprocal sensitivity ($1/\eta$) for dark spins DS2+ and DS1 (See methods). $\ket{\psi_2}$ achieves an optimal sensitivity, surpassing the NV1 single-spin performance by 5.28 dB for detecting DS2+, while showing a sensitivity degradation of -0.98 dB for detecting DS1. The squares (solid lines) represent experimental results (predictions). \textbf{F.} Sensitivity gain of the entangled state sensor compared to NV1 with the same sensing duration $\mathrm{t_{DEER}}$.}

	\label{Fig3}
\end{figure*}

This short separation significantly reduces the overhead associated with entangled state preparation and detection. With a primary time cost of $\tau_1=1/(8A_{\mathrm{NV1,NV2}})\sim 0.2\ \mu$s for the interaction, we prepared two types of entangled states, $\ket{\psi_{1}}=\left(\ket{\Uparrow\Uparrow}+i\ket{\Downarrow\Downarrow}\right)/\sqrt{2}, \ket{\psi_{2}}=\left(\ket{\Uparrow\Downarrow}+i\ket{\Downarrow\Uparrow}\right)/\sqrt{2}$, using the sequence shown in Fig.\ref{Fig2}B, where $\Uparrow(\Downarrow)$ denotes $m_s=+1(-1)$  for spin-1 system. These two states exhibit complementary sensitivities to the surface electronic spin bath (Fig.\ref{Fig2}D). 
As a result, we observe an enhanced coherence time of $21.9\  \mu s$ for state $\ket{\psi_{2}}$ and a reduced coherence time of $4.8\  \mu s$ for state $\ket{\psi_{1}}$.
These results starkly contrast with previous entangled NV pairs, which exhibit identical coherence times due to a large separation of 25 nm \cite{Joerg_Entanglement_2013}.

We employ $\psi_{2}$ based DEER spectroscopy to achieve entangled nanoscale sensing of dark spins (Fig.\ref{Fig2}B). When the frequency of the target spin's  $\pi$-flip pulse matches its resonance frequency, the interaction between target spins and $\psi_{2}$ remains coupled, inducing a detectable phase shift in $\psi_{2}$ during disentanglement. 
Frequency scanning reveals three distinct resonance peaks  at 0.0854, 0.2495, and 0.6940 GHz (Fig.\ref{Fig2}E), corresponding to individual dark electron spins. To verify single-spin origin, we measure each dark spin-induced phase dynamics as a function of the evolution time t. The persistence of coherent monotonic oscillations at all the resonance frequencies (Fig.\ref{Fig3}A) unambiguously confirms single-spin signatures, as ensemble spins would average the oscillation and yield decoherence-like signals. The oscillation frequency (Fig.\ref{Fig3}B) is governed by the differential dipole-dipole coupling between the dark spin and the two NV centers.

Nanoscale spin sensing enables simultaneous identification and spatial mapping of dark spins through their unique magnetic fingerprints. By analyzing their magnetic field-dependent energy splitting (Fig.\ref{Fig3}C, Extended Data), we identify distinct spin configurations: DS1 exhibits characteristic spin-1/2 behavior with isotropic Zeeman splitting with susceptibility of $2.8$ MHz/Gauss across different orientations, while DS2 displays spin triplet characteristics featuring zero-field splitting with axial term D = 0.2966 GHz and non-axial term E = 0.3003 GHz (SI), manifested as two separate peaks at zero magnetic field and nonlinear energy-field dependence at finite magnetic fields.  

Leveraging the anisotropic nature of dipole-dipole interactions, we establish a protocol for nanoscale spatial mapping of dark spins. This approach capitalizes on the interaction's dual dependence on (1) inter-spin distance ($r^{-3}$ scaling) and (2) orientation relative to the applied vector magnetic field $\boldsymbol{\mathrm{B_{0}}}$. By systematically varying $\boldsymbol{\mathrm{B_{0}}}$ and measuring resultant coupling strengths, we reconstruct the three-dimensional spin distribution with $\sim$ 0.3 nm positional uncertainty (Extended Data). The reconstructed map (Fig.\ref{Fig3}D) reveals a clustered spin arrangement with a characteristic spacing of $\sim$ 3.0 nm, and these two dark spins are located inside the diamond crystal.

The entangled sensor state $\psi_2$ exhibits strongly contrasting responses to DS1 and DS2 (Fig.\ref{Fig3}E), arising from their distinct spatial configurations relative to the NV pair. This spatial selectivity enables simultaneous enhancement of the DS2 coupling strength and suppression of DS1 signals. We quantify the sensitivity by modeling the dipole-dipole interaction as an effective magnetic field (Methods). 
Quantitative analysis shows that $\psi_2$ provides an optimal 5.28 dB sensitivity enhancement over single-NV detection for DS2 while simultaneously suppressing unwanted DS1 signals by -0.98 dB. These results establish entanglement engineering as a powerful strategy for nanoscale spin discrimination, particularly in dense spin environments where traditional detection methods may fail.

\begin{figure*} [ht]
	\begin{center}
	\includegraphics[scale=1]{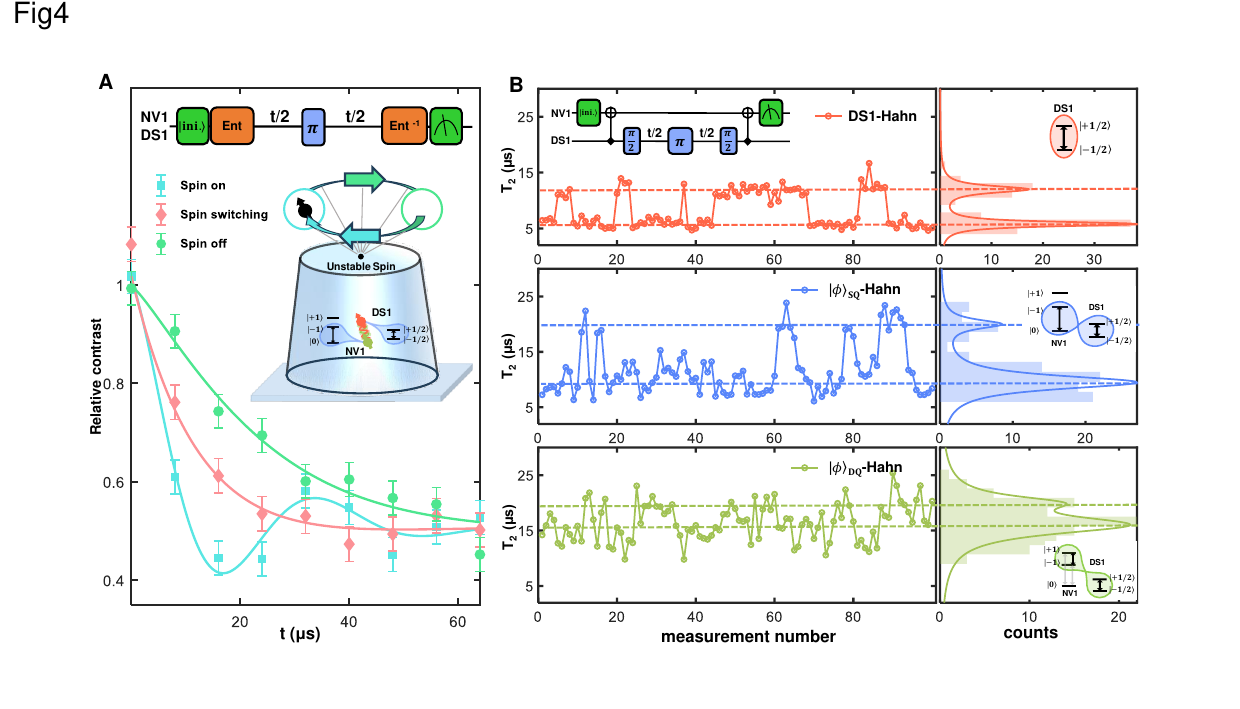}
	\end{center}
	\caption{Unstable dark spins detection using entangled sensors. \textbf{A.} Three typical results for detecting an unstable dark spin in three states using NV1-DS1-entangled state $\ket{\phi}_\mathrm{SQ}$. State-1 (Spin on): The spin remains stable during the DEER measurement and induces a clear oscillation signature (square). State-2 (Spin off): The spin disappears during the measurement, resulting in a pure-decay (circular) with the same decoherence rate as in State-1. State-3 (Spin switching): The spin switches between two spin-states during measurement, causing an additional decoherence (rhombus). Each experimental point spends a measurement time of 180 seconds. Solid lines represent the fitting curve. The DS1 and interficial dark spin share the same $\pi$ pulse in the inset sequence. 
    \textbf{B.} Real-time $T_2$ tracks (Left) and their statistics (Right) for three different spin states: $\frac{1}{\sqrt{2}}\ket{(\ket{\downarrow}+\ket{\uparrow})}_\mathrm{DS1}$, $\ket{\phi}_\mathrm{SQ}$, and $\ket{\phi}_\mathrm{DQ}$.  
    }
	\label{Fig4}
\end{figure*}

Building on established techniques that employ isolated dark spins as auxiliary sensors with extended sensing range \cite{Lukin_reporter_2014,Ungar2024}, we integrate this approach with our entangled nanoscale sensing protocol to achieve single-spin detection. 
Entangled NV1-DS1 sensor pairs are explored to detect the interficial spins for their proximal location near the diamond-air interface.  
The experiment sequence begins with a SWAP-gate-mediated polarization transfer to DS1, followed by preparation of an entangled state $\ket{\phi}_{\mathrm{SQ}}=(\ket{\Downarrow, \uparrow}-i\ket{0, \downarrow})/\sqrt{2}$ between NV1 ($m_s=0,-1$) and DS1, and finally proceeds to implementation of interface-sensitive DEER measurements targeting dominant spin-1/2 species (Fig.\ref{Fig4}A).
The measurements reveal a metastable dark spin exhibiting stochastic switching between spin-1/2 and spin-0 configurations. When in the spin-1/2 state, we observe clear coherent oscillations, while these oscillations completely vanish during spin-0 intervals. The measured coupling strength of 27(7) kHz (corresponding to an estimated 11 nm above the pair) and switching dynamics suggest that this spin is located within the interfacial charge trapping layer \cite{Hollenberg_Surface_2019}, where local electrical environments likely drive its metastable behavior.  

Metastable spins play a dual role in quantum sensing applications, functioning as sensitive nanoscale probes and potential noise sources. These dynamic spin systems facilitate the investigation of rapid chemical processes \cite{Murai2003,Huang2025} and interfacial charge trapping dynamics \cite{Jayich_charge_2019}, where our entanglement protocol can enhance the coupling strength, thereby reducing the detection time for fast sampling. However, they simultaneously introduce significant decoherence channels (Fig.\ref{Fig4}B) that degrade the sensitivity of detecting stable target spins. To address this, we engineer the entangled states $\ket{\phi}_{\mathrm{DQ}}=(\ket{\Downarrow, \uparrow}-i\ket{\Uparrow, \downarrow})/\sqrt{2}$, which demonstrates remarkable noise resilience by extending coherence times from 6 $\mu s$ (DS1 alone) to 15 $\mu s$. 
This 2.5 times improvement stems from enhanced coupling cancellation, as the NV1-metastable spin coupling doubles in $\ket{\phi}_{\mathrm{DQ}}$, approaching the coupling strength between DS1 and the target at relatively short distances (Fig.\ref {Fig4}A inset). 
This dual functionality offers new avenues for studying complex quantum systems where static and dynamic spin species coexist.

\section{Discussion and outlook}

The emergence of diverse atomic and molecular systems as nanoscale quantum sensors has provided the toolset for revolutionizing structural analysis and microscopic investigation with unprecedented precision and spatial resolution. 
The present work has demonstrated a universal and scalable approach to enhance nanoscale sensing by harnessing quantum entanglement in realistic noise environments. A central achievement of our work lies in developing a robust framework that maintains the full benefits of quantum entanglement while effectively addressing its fundamental challenges. Through the dedicated engineering of quantum sensor arrays to form large-scale entangled systems, coupled with high-fidelity quantum control and readout protocols, our methodology pushes the boundaries of nanoscale quantum sensing toward its fundamental physical limits. 

Another immediate application of this technology involves developing entangled quantum sensors into functional scanning probe systems capable of deterministic specimen positioning at optimized locations. This advancement would dramatically enhance both measurement precision and operational reliability for various nanoscale quantum sensing applications.

\section{Methods}

\subsection{Diamond sample fabrication}
The diamond substrate used in the experiment is a 50 $\ mum$-thick (100)-oriented diamond with an isotopically enriched (99.999\% $^{12}$C) epitaxially grown layer. The NV pairs in diamond pillars are fabricated via a self-aligned patterning technique, where 15-nm-radius mask holes in a PMMA layer confine the NVs in close proximity and in the centra region of diamond pillars. These NV centers are generated by implanting 15 keV $^{15}\mathrm{N}^{+}$ ions, with the ion dose controlled to ensure approximately 10 ions are implanted per mask hole.



\subsection{Spatial resolution}
The spatial resolution of our NV center-based quantum sensor is fundamentally determined by its effective sensing area for surface dark electron spins, following the established effective sensing volume methodology \cite{Staudacher2013}. For a dense layer of surface spins interacting via dipolar coupling $A_\mathrm{eff,i}$, we first index all spins in descending order of their coupling strengths $A_\mathrm{eff,1} \geq A_\mathrm{eff,2} \geq ...$. The cumulative signal contribution is quantified as $S(N) = \sqrt{\sum_{i=1}^N A^2_{\mathrm{eff},i}}$, from which we define the effective sensing area as the spatial extent containing the minimum number of spins ($N_\mathrm{eff}$) that contribute $70\%$ of the total detectable signal $S(N=N_\mathrm{eff})/S(N=\inf)= 70\% $. A single spin within this region can be effectively distinguished from other spins.

\subsection{Spin-detection Sensitivity}
We quantify the sensitivity of target spins detection through dipole-dipole interaction, treating these interactions as an effective magnetic field.
For a DEER detection sequence with the sensing time ($t=t_\mathrm{DEER}$), the sensitivity can be expressed as \cite{Barry2020}
\begin{equation*}
\eta= \frac{1}{\pi A_\mathrm{eff}\sqrt{\tau}e^{-\left(t_\mathrm{DEER} / T_2\right)^p}}\sqrt{1+\frac{1}{C^2n_\mathrm{avg}}}\sqrt{1+\frac{t_\mathrm{ext}}{t_\mathrm{DEER}}}
\label{eq_sen}
\end{equation*}
, where $T_2$ is the coherence time and $p$ is the stretched exponential parameter characterizing local spin environments. $n_\mathrm{avg}$ denotes the average photon count per readout (experimentally determined as $n_\mathrm{avg} = 0.12$ for $\ket{\psi_2}$ and $n_\mathrm{avg} = 0.06$ for a single NV center), and $C$ is the measurement contrast. Furthermore, the optical initialization time, readout time, and other external times are represented by $ t_\mathrm{ext}=t_I+t_R+t_{\mathrm{other}}$. Complete experimental parameters and detailed derivations are provided in the Supplementary Information.

\subsection{Spin localization}
The spatial positioning of target spins is achieved through precise characterization of their magnetic dipole interactions. For spin-1/2 systems (e.g., DS1), the dipole moment aligns directly with the external magnetic field $B_{ext}$. However, for spin-1 systems (NV1, NV2, DS2), the dipole orientation is determined by the interplay between their intrinsic anisotropy axes and $B_{ext}$. Given their distinct principal axis orientations, we first define the dipole moment in each spin's principal coordinate system ($[\mathbf{D_{i}^x},\mathbf{D_{i}^y},\mathbf{D_{i}^z},]$),
\begin{equation*}
\Phi_j^i=\langle\phi_j^i|S_x|\phi_j^i\rangle \mathbf{D_{i}^x}+\langle\phi_j^i|S_y|\phi_j^i\rangle \mathbf{D_{i}^y}+\langle\phi_j^i|S_z|\phi_j^i\rangle \mathbf{D_{i}^z}
\label{eq_Loc1}
\end{equation*}
, where $\phi_i^j$ represents the j-th eigenstate of the spin's Hamiltonian under $B_{ext}$. For coupled spin-1 pairs, the interaction strength between states $\Phi_{j_1}^i$ and$\Phi_{j_2}^i$ follows the dipolar coupling formula
\begin{equation*}
\nu_{j_1,j_2}\propto\frac{a}{r^3}\left[\Phi^i_{j_1} \Phi^i_{j_2}-3\frac{(\Phi^i_{j_1}\cdot \mathbf{r})(\Phi^i_{j_2}\cdot \mathbf{r}))}{r^2}\right].
\label{eq_Loc2}
\end{equation*}
The experimentally measurable coupling strength for transitions $\Phi_{j_1}^i \rightarrow \Phi_{j'_1}^i$ (sensor) and $\Phi_{j_2}^i \rightarrow \Phi_{j'_2}^i$ (target) is
\begin{equation*}
A_{j_1\rightarrow j'_1,j_2\rightarrow j'_2}=(\nu_{j_1,j_2}-\nu_{j_1,j'_2})-(\nu_{j'_1,j_2}-\nu_{j'_1,j'_2}).
\label{eq_Loc3}
\end{equation*}

The localization method then involves: (1) measuring coupling strengths under varying $B_\mathrm{ext}$; (2) numerically searching for the position vector $\mathbf{r}$ that best reproduces all observed couplings; (3) determining the intersection of solutions from multiple field orientations.

\section{Acknowledgements}
Funding: This work was supported by the National Natural Science Foundation of China (grants no. T2325023, T2388102, 92265204, 12104446, 12104447, and 12261160569) and the Innovation Program for Quantum Science and Technology (grant no. 2021ZD0302200). 

Author contributions: Y.W. conceived the idea. Y.W., M.W., and X.Z. designed the experiment. M.W. and X.Y. prepared the sample. X.Z., Y.G., H.S., Z.C., and W.J. performed the experiments and analyzed the data. Y.W., X.Z., M.W., K.X., F.S., and J.D. wrote the manuscript. All authors discussed the results and commented on the manuscript. J.D. and Y.W. supervised the project.  

Competing interests: The authors declare that they have no competing interests. 

Data and materials availability: All data needed to evaluate the conclusions in the paper are present in the paper and/or the Supplementary Materials.
\bibliography{references}

\begin{thebibliography}{46}%
\makeatletter
\providecommand \@ifxundefined [1]{%
 \@ifx{#1\undefined}
}%
\providecommand \@ifnum [1]{%
 \ifnum #1\expandafter \@firstoftwo
 \else \expandafter \@secondoftwo
 \fi
}%
\providecommand \@ifx [1]{%
 \ifx #1\expandafter \@firstoftwo
 \else \expandafter \@secondoftwo
 \fi
}%
\providecommand \natexlab [1]{#1}%
\providecommand \enquote  [1]{``#1''}%
\providecommand \bibnamefont  [1]{#1}%
\providecommand \bibfnamefont [1]{#1}%
\providecommand \citenamefont [1]{#1}%
\providecommand \href@noop [0]{\@secondoftwo}%
\providecommand \href [0]{\begingroup \@sanitize@url \@href}%
\providecommand \@href[1]{\@@startlink{#1}\@@href}%
\providecommand \@@href[1]{\endgroup#1\@@endlink}%
\providecommand \@sanitize@url [0]{\catcode `\\12\catcode `\$12\catcode
  `\&12\catcode `\#12\catcode `\^12\catcode `\_12\catcode `\%12\relax}%
\providecommand \@@startlink[1]{}%
\providecommand \@@endlink[0]{}%
\providecommand \url  [0]{\begingroup\@sanitize@url \@url }%
\providecommand \@url [1]{\endgroup\@href {#1}{\urlprefix }}%
\providecommand \urlprefix  [0]{URL }%
\providecommand \Eprint [0]{\href }%
\providecommand \doibase [0]{https://doi.org/}%
\providecommand \selectlanguage [0]{\@gobble}%
\providecommand \bibinfo  [0]{\@secondoftwo}%
\providecommand \bibfield  [0]{\@secondoftwo}%
\providecommand \translation [1]{[#1]}%
\providecommand \BibitemOpen [0]{}%
\providecommand \bibitemStop [0]{}%
\providecommand \bibitemNoStop [0]{.\EOS\space}%
\providecommand \EOS [0]{\spacefactor3000\relax}%
\providecommand \BibitemShut  [1]{\csname bibitem#1\endcsname}%
\let\auto@bib@innerbib\@empty
\bibitem [{\citenamefont {Du}\ \emph {et~al.}(2024)\citenamefont {Du},
  \citenamefont {Shi}, \citenamefont {Kong}, \citenamefont {Jelezko},\ and\
  \citenamefont {Wrachtrup}}]{RMP_Du_2024}%
  \BibitemOpen
  \bibfield  {author} {\bibinfo {author} {\bibfnamefont {J.}~\bibnamefont
  {Du}}, \bibinfo {author} {\bibfnamefont {F.}~\bibnamefont {Shi}}, \bibinfo
  {author} {\bibfnamefont {X.}~\bibnamefont {Kong}}, \bibinfo {author}
  {\bibfnamefont {F.}~\bibnamefont {Jelezko}},\ and\ \bibinfo {author}
  {\bibfnamefont {J.}~\bibnamefont {Wrachtrup}},\ }\bibfield  {title} {\bibinfo
  {title} {Single-molecule scale magnetic resonance spectroscopy using quantum
  diamond sensors},\ }\href {https://doi.org/10.1103/RevModPhys.96.025001}
  {\bibfield  {journal} {\bibinfo  {journal} {Rev. Mod. Phys.}\ }\textbf
  {\bibinfo {volume} {96}},\ \bibinfo {pages} {025001} (\bibinfo {year}
  {2024})}\BibitemShut {NoStop}%
\bibitem [{\citenamefont {Boss}\ \emph {et~al.}(2017)\citenamefont {Boss},
  \citenamefont {Cujia}, \citenamefont {Zopes},\ and\ \citenamefont
  {Degen}}]{Degen_resolution_2017}%
  \BibitemOpen
  \bibfield  {author} {\bibinfo {author} {\bibfnamefont {J.~M.}\ \bibnamefont
  {Boss}}, \bibinfo {author} {\bibfnamefont {K.~S.}\ \bibnamefont {Cujia}},
  \bibinfo {author} {\bibfnamefont {J.}~\bibnamefont {Zopes}},\ and\ \bibinfo
  {author} {\bibfnamefont {C.~L.}\ \bibnamefont {Degen}},\ }\bibfield  {title}
  {\bibinfo {title} {Quantum sensing with arbitrary frequency resolution},\
  }\href {https://doi.org/10.1126/science.aam7009} {\bibfield  {journal}
  {\bibinfo  {journal} {Science}\ }\textbf {\bibinfo {volume} {356}},\ \bibinfo
  {pages} {837} (\bibinfo {year} {2017})}\BibitemShut {NoStop}%
\bibitem [{\citenamefont {Schmitt}\ \emph {et~al.}(2017)\citenamefont
  {Schmitt}, \citenamefont {Gefen}, \citenamefont {Sturner}, \citenamefont
  {Unden}, \citenamefont {Wolff}, \citenamefont {Muller}, \citenamefont
  {Scheuer}, \citenamefont {Naydenov}, \citenamefont {Markham}, \citenamefont
  {Pezzagna}, \citenamefont {Meijer}, \citenamefont {Schwarz}, \citenamefont
  {Plenio}, \citenamefont {Retzker}, \citenamefont {McGuinness},\ and\
  \citenamefont {Jelezko}}]{Jelezko_submillihertz_2017}%
  \BibitemOpen
  \bibfield  {author} {\bibinfo {author} {\bibfnamefont {S.}~\bibnamefont
  {Schmitt}}, \bibinfo {author} {\bibfnamefont {T.}~\bibnamefont {Gefen}},
  \bibinfo {author} {\bibfnamefont {F.~M.}\ \bibnamefont {Sturner}}, \bibinfo
  {author} {\bibfnamefont {T.}~\bibnamefont {Unden}}, \bibinfo {author}
  {\bibfnamefont {G.}~\bibnamefont {Wolff}}, \bibinfo {author} {\bibfnamefont
  {C.}~\bibnamefont {Muller}}, \bibinfo {author} {\bibfnamefont
  {J.}~\bibnamefont {Scheuer}}, \bibinfo {author} {\bibfnamefont
  {B.}~\bibnamefont {Naydenov}}, \bibinfo {author} {\bibfnamefont
  {M.}~\bibnamefont {Markham}}, \bibinfo {author} {\bibfnamefont
  {S.}~\bibnamefont {Pezzagna}}, \bibinfo {author} {\bibfnamefont
  {J.}~\bibnamefont {Meijer}}, \bibinfo {author} {\bibfnamefont
  {I.}~\bibnamefont {Schwarz}}, \bibinfo {author} {\bibfnamefont
  {M.}~\bibnamefont {Plenio}}, \bibinfo {author} {\bibfnamefont
  {A.}~\bibnamefont {Retzker}}, \bibinfo {author} {\bibfnamefont {L.~P.}\
  \bibnamefont {McGuinness}},\ and\ \bibinfo {author} {\bibfnamefont
  {F.}~\bibnamefont {Jelezko}},\ }\bibfield  {title} {\bibinfo {title}
  {Submillihertz magnetic spectroscopy performed with a nanoscale quantum
  sensor},\ }\href {https://doi.org/10.1126/science.aam5532} {\bibfield
  {journal} {\bibinfo  {journal} {Science}\ }\textbf {\bibinfo {volume}
  {356}},\ \bibinfo {pages} {832} (\bibinfo {year} {2017})}\BibitemShut
  {NoStop}%
\bibitem [{\citenamefont {Aslam}\ \emph {et~al.}(2017)\citenamefont {Aslam},
  \citenamefont {Pfender}, \citenamefont {Neumann}, \citenamefont {Reuter},
  \citenamefont {Zappe}, \citenamefont {Fávaro~de Oliveira}, \citenamefont
  {Denisenko}, \citenamefont {Sumiya}, \citenamefont {Onoda}, \citenamefont
  {Isoya},\ and\ \citenamefont {Wrachtrup}}]{Joerg_Chemical_2017}%
  \BibitemOpen
  \bibfield  {author} {\bibinfo {author} {\bibfnamefont {N.}~\bibnamefont
  {Aslam}}, \bibinfo {author} {\bibfnamefont {M.}~\bibnamefont {Pfender}},
  \bibinfo {author} {\bibfnamefont {P.}~\bibnamefont {Neumann}}, \bibinfo
  {author} {\bibfnamefont {R.}~\bibnamefont {Reuter}}, \bibinfo {author}
  {\bibfnamefont {A.}~\bibnamefont {Zappe}}, \bibinfo {author} {\bibfnamefont
  {F.}~\bibnamefont {Fávaro~de Oliveira}}, \bibinfo {author} {\bibfnamefont
  {A.}~\bibnamefont {Denisenko}}, \bibinfo {author} {\bibfnamefont
  {H.}~\bibnamefont {Sumiya}}, \bibinfo {author} {\bibfnamefont
  {S.}~\bibnamefont {Onoda}}, \bibinfo {author} {\bibfnamefont
  {J.}~\bibnamefont {Isoya}},\ and\ \bibinfo {author} {\bibfnamefont
  {J.}~\bibnamefont {Wrachtrup}},\ }\bibfield  {title} {\bibinfo {title}
  {Nanoscale nuclear magnetic resonance with chemical resolution},\ }\href
  {https://doi.org/10.1126/science.aam8697} {\bibfield  {journal} {\bibinfo
  {journal} {Science}\ }\textbf {\bibinfo {volume} {357}},\ \bibinfo {pages}
  {67} (\bibinfo {year} {2017})}\BibitemShut {NoStop}%
\bibitem [{\citenamefont {Casola}\ \emph {et~al.}(2018)\citenamefont {Casola},
  \citenamefont {van~der Sar},\ and\ \citenamefont {Yacoby}}]{Casola2018}%
  \BibitemOpen
  \bibfield  {author} {\bibinfo {author} {\bibfnamefont {F.}~\bibnamefont
  {Casola}}, \bibinfo {author} {\bibfnamefont {T.}~\bibnamefont {van~der
  Sar}},\ and\ \bibinfo {author} {\bibfnamefont {A.}~\bibnamefont {Yacoby}},\
  }\bibfield  {title} {\bibinfo {title} {Probing condensed matter physics with
  magnetometry based on nitrogen-vacancy centres in diamond},\ }\bibfield
  {journal} {\bibinfo  {journal} {Nature Reviews Materials}\ }\textbf {\bibinfo
  {volume} {3}},\ \href {https://doi.org/10.1038/natrevmats.2017.88}
  {10.1038/natrevmats.2017.88} (\bibinfo {year} {2018})\BibitemShut {NoStop}%
\bibitem [{\citenamefont {Abobeih}\ \emph {et~al.}(2019)\citenamefont
  {Abobeih}, \citenamefont {Randall}, \citenamefont {Bradley}, \citenamefont
  {Bartling}, \citenamefont {Bakker}, \citenamefont {Degen}, \citenamefont
  {Markham}, \citenamefont {Twitchen},\ and\ \citenamefont
  {Taminiau}}]{Tim_Nspin_cluster_2019}%
  \BibitemOpen
  \bibfield  {author} {\bibinfo {author} {\bibfnamefont {M.~H.}\ \bibnamefont
  {Abobeih}}, \bibinfo {author} {\bibfnamefont {J.}~\bibnamefont {Randall}},
  \bibinfo {author} {\bibfnamefont {C.~E.}\ \bibnamefont {Bradley}}, \bibinfo
  {author} {\bibfnamefont {H.~P.}\ \bibnamefont {Bartling}}, \bibinfo {author}
  {\bibfnamefont {M.~A.}\ \bibnamefont {Bakker}}, \bibinfo {author}
  {\bibfnamefont {M.~J.}\ \bibnamefont {Degen}}, \bibinfo {author}
  {\bibfnamefont {M.}~\bibnamefont {Markham}}, \bibinfo {author} {\bibfnamefont
  {D.~J.}\ \bibnamefont {Twitchen}},\ and\ \bibinfo {author} {\bibfnamefont
  {T.~H.}\ \bibnamefont {Taminiau}},\ }\bibfield  {title} {\bibinfo {title}
  {Atomic-scale imaging of a 27-nuclear-spin cluster using a quantum sensor},\
  }\href {https://doi.org/10.1038/s41586-019-1834-7} {\bibfield  {journal}
  {\bibinfo  {journal} {Nature}\ }\textbf {\bibinfo {volume} {576}},\ \bibinfo
  {pages} {411} (\bibinfo {year} {2019})}\BibitemShut {NoStop}%
\bibitem [{\citenamefont {Song}\ \emph {et~al.}(2021)\citenamefont {Song},
  \citenamefont {Sun}, \citenamefont {Anderson}, \citenamefont {Wang},
  \citenamefont {Qian}, \citenamefont {Taniguchi}, \citenamefont {Watanabe},
  \citenamefont {McGuire}, \citenamefont {Stohr}, \citenamefont {Xiao},
  \citenamefont {Cao}, \citenamefont {Wrachtrup},\ and\ \citenamefont
  {Xu}}]{Xu_moire_2021}%
  \BibitemOpen
  \bibfield  {author} {\bibinfo {author} {\bibfnamefont {T.}~\bibnamefont
  {Song}}, \bibinfo {author} {\bibfnamefont {Q.~C.}\ \bibnamefont {Sun}},
  \bibinfo {author} {\bibfnamefont {E.}~\bibnamefont {Anderson}}, \bibinfo
  {author} {\bibfnamefont {C.}~\bibnamefont {Wang}}, \bibinfo {author}
  {\bibfnamefont {J.}~\bibnamefont {Qian}}, \bibinfo {author} {\bibfnamefont
  {T.}~\bibnamefont {Taniguchi}}, \bibinfo {author} {\bibfnamefont
  {K.}~\bibnamefont {Watanabe}}, \bibinfo {author} {\bibfnamefont {M.~A.}\
  \bibnamefont {McGuire}}, \bibinfo {author} {\bibfnamefont {R.}~\bibnamefont
  {Stohr}}, \bibinfo {author} {\bibfnamefont {D.}~\bibnamefont {Xiao}},
  \bibinfo {author} {\bibfnamefont {T.}~\bibnamefont {Cao}}, \bibinfo {author}
  {\bibfnamefont {J.}~\bibnamefont {Wrachtrup}},\ and\ \bibinfo {author}
  {\bibfnamefont {X.}~\bibnamefont {Xu}},\ }\bibfield  {title} {\bibinfo
  {title} {Direct visualization of magnetic domains and moire magnetism in
  twisted 2d magnets},\ }\href {https://doi.org/10.1126/science.abj7478}
  {\bibfield  {journal} {\bibinfo  {journal} {Science}\ }\textbf {\bibinfo
  {volume} {374}},\ \bibinfo {pages} {1140} (\bibinfo {year}
  {2021})}\BibitemShut {NoStop}%
\bibitem [{\citenamefont {Borst}\ \emph {et~al.}(2023)\citenamefont {Borst},
  \citenamefont {Vree}, \citenamefont {Lowther}, \citenamefont {Teepe},
  \citenamefont {Kurdi}, \citenamefont {Bertelli}, \citenamefont {Simon},
  \citenamefont {Blanter},\ and\ \citenamefont {van~der
  Sar}}]{Van_Spinwave_2023}%
  \BibitemOpen
  \bibfield  {author} {\bibinfo {author} {\bibfnamefont {M.}~\bibnamefont
  {Borst}}, \bibinfo {author} {\bibfnamefont {P.~H.}\ \bibnamefont {Vree}},
  \bibinfo {author} {\bibfnamefont {A.}~\bibnamefont {Lowther}}, \bibinfo
  {author} {\bibfnamefont {A.}~\bibnamefont {Teepe}}, \bibinfo {author}
  {\bibfnamefont {S.}~\bibnamefont {Kurdi}}, \bibinfo {author} {\bibfnamefont
  {I.}~\bibnamefont {Bertelli}}, \bibinfo {author} {\bibfnamefont {B.~G.}\
  \bibnamefont {Simon}}, \bibinfo {author} {\bibfnamefont {Y.~M.}\ \bibnamefont
  {Blanter}},\ and\ \bibinfo {author} {\bibfnamefont {T.}~\bibnamefont {van~der
  Sar}},\ }\bibfield  {title} {\bibinfo {title} {Observation and control of
  hybrid spin-wave-meissner-current transport modes},\ }\href
  {https://doi.org/10.1126/science.adj7576} {\bibfield  {journal} {\bibinfo
  {journal} {Science}\ }\textbf {\bibinfo {volume} {382}},\ \bibinfo {pages}
  {430} (\bibinfo {year} {2023})}\BibitemShut {NoStop}%
\bibitem [{\citenamefont {Ji}\ \emph {et~al.}(2024)\citenamefont {Ji},
  \citenamefont {Liu}, \citenamefont {Guo}, \citenamefont {Hu}, \citenamefont
  {Zhou}, \citenamefont {Dai}, \citenamefont {Chen}, \citenamefont {Yu},
  \citenamefont {Wang}, \citenamefont {Xia}, \citenamefont {Shi}, \citenamefont
  {Wang},\ and\ \citenamefont {Du}}]{Ya_Correlated_2024}%
  \BibitemOpen
  \bibfield  {author} {\bibinfo {author} {\bibfnamefont {W.~T.}\ \bibnamefont
  {Ji}}, \bibinfo {author} {\bibfnamefont {Z.~X.}\ \bibnamefont {Liu}},
  \bibinfo {author} {\bibfnamefont {Y.~H.}\ \bibnamefont {Guo}}, \bibinfo
  {author} {\bibfnamefont {Z.~H.}\ \bibnamefont {Hu}}, \bibinfo {author}
  {\bibfnamefont {J.~Y.}\ \bibnamefont {Zhou}}, \bibinfo {author}
  {\bibfnamefont {S.~H.}\ \bibnamefont {Dai}}, \bibinfo {author} {\bibfnamefont
  {Y.}~\bibnamefont {Chen}}, \bibinfo {author} {\bibfnamefont {P.}~\bibnamefont
  {Yu}}, \bibinfo {author} {\bibfnamefont {M.~Q.}\ \bibnamefont {Wang}},
  \bibinfo {author} {\bibfnamefont {K.~W.}\ \bibnamefont {Xia}}, \bibinfo
  {author} {\bibfnamefont {F.~Z.}\ \bibnamefont {Shi}}, \bibinfo {author}
  {\bibfnamefont {Y.}~\bibnamefont {Wang}},\ and\ \bibinfo {author}
  {\bibfnamefont {J.~F.}\ \bibnamefont {Du}},\ }\bibfield  {title} {\bibinfo
  {title} {Correlated sensing with a solid-state quantum multisensor system for
  atomic-scale structural analysis},\ }\bibfield  {journal} {\bibinfo
  {journal} {Nature Photonics}\ }\textbf {\bibinfo {volume} {18}},\ \href
  {https://doi.org/10.1038/s41566-023-01352-4} {10.1038/s41566-023-01352-4}
  (\bibinfo {year} {2024})\BibitemShut {NoStop}%
\bibitem [{\citenamefont {Rovny}\ \emph {et~al.}(2024)\citenamefont {Rovny},
  \citenamefont {Gopalakrishnan}, \citenamefont {Jayich}, \citenamefont
  {Maletinsky}, \citenamefont {Demler},\ and\ \citenamefont
  {de~Leon}}]{Many-body_NRP_2024}%
  \BibitemOpen
  \bibfield  {author} {\bibinfo {author} {\bibfnamefont {J.}~\bibnamefont
  {Rovny}}, \bibinfo {author} {\bibfnamefont {S.}~\bibnamefont
  {Gopalakrishnan}}, \bibinfo {author} {\bibfnamefont {A.~C.~B.}\ \bibnamefont
  {Jayich}}, \bibinfo {author} {\bibfnamefont {P.}~\bibnamefont {Maletinsky}},
  \bibinfo {author} {\bibfnamefont {E.}~\bibnamefont {Demler}},\ and\ \bibinfo
  {author} {\bibfnamefont {N.~P.}\ \bibnamefont {de~Leon}},\ }\bibfield
  {title} {\bibinfo {title} {Nanoscale diamond quantum sensors for many-body
  physics},\ }\href {https://doi.org/10.1038/s42254-024-00775-4} {\bibfield
  {journal} {\bibinfo  {journal} {Nature Reviews Physics}\ }\textbf {\bibinfo
  {volume} {6}},\ \bibinfo {pages} {753} (\bibinfo {year} {2024})}\BibitemShut
  {NoStop}%
\bibitem [{\citenamefont {Andersen}\ \emph {et~al.}(2019)\citenamefont
  {Andersen}, \citenamefont {Dwyer}, \citenamefont {Sanchez-Yamagishi},
  \citenamefont {Rodriguez-Nieva}, \citenamefont {Agarwal}, \citenamefont
  {Watanabe}, \citenamefont {Taniguchi}, \citenamefont {Demler}, \citenamefont
  {Kim}, \citenamefont {Park},\ and\ \citenamefont
  {Lukin}}]{Lukin_Graphene_2019}%
  \BibitemOpen
  \bibfield  {author} {\bibinfo {author} {\bibfnamefont {T.~I.}\ \bibnamefont
  {Andersen}}, \bibinfo {author} {\bibfnamefont {B.~L.}\ \bibnamefont {Dwyer}},
  \bibinfo {author} {\bibfnamefont {J.~D.}\ \bibnamefont {Sanchez-Yamagishi}},
  \bibinfo {author} {\bibfnamefont {J.~F.}\ \bibnamefont {Rodriguez-Nieva}},
  \bibinfo {author} {\bibfnamefont {K.}~\bibnamefont {Agarwal}}, \bibinfo
  {author} {\bibfnamefont {K.}~\bibnamefont {Watanabe}}, \bibinfo {author}
  {\bibfnamefont {T.}~\bibnamefont {Taniguchi}}, \bibinfo {author}
  {\bibfnamefont {E.~A.}\ \bibnamefont {Demler}}, \bibinfo {author}
  {\bibfnamefont {P.}~\bibnamefont {Kim}}, \bibinfo {author} {\bibfnamefont
  {H.}~\bibnamefont {Park}},\ and\ \bibinfo {author} {\bibfnamefont {M.~D.}\
  \bibnamefont {Lukin}},\ }\bibfield  {title} {\bibinfo {title}
  {Electron-phonon instability in graphene revealed by global and local noise
  probes},\ }\href {https://doi.org/10.1126/science.aaw2104} {\bibfield
  {journal} {\bibinfo  {journal} {Science}\ }\textbf {\bibinfo {volume}
  {364}},\ \bibinfo {pages} {154} (\bibinfo {year} {2019})}\BibitemShut
  {NoStop}%
\bibitem [{\citenamefont {Zu}\ \emph {et~al.}(2021)\citenamefont {Zu},
  \citenamefont {Machado}, \citenamefont {Ye}, \citenamefont {Choi},
  \citenamefont {Kobrin}, \citenamefont {Mittiga}, \citenamefont {Hsieh},
  \citenamefont {Bhattacharyya}, \citenamefont {Markham}, \citenamefont
  {Twitchen}, \citenamefont {Jarmola}, \citenamefont {Budker}, \citenamefont
  {Laumann}, \citenamefont {Moore},\ and\ \citenamefont
  {Yao}}]{Norman_Hydrodynamics_2021}%
  \BibitemOpen
  \bibfield  {author} {\bibinfo {author} {\bibfnamefont {C.}~\bibnamefont
  {Zu}}, \bibinfo {author} {\bibfnamefont {F.}~\bibnamefont {Machado}},
  \bibinfo {author} {\bibfnamefont {B.}~\bibnamefont {Ye}}, \bibinfo {author}
  {\bibfnamefont {S.}~\bibnamefont {Choi}}, \bibinfo {author} {\bibfnamefont
  {B.}~\bibnamefont {Kobrin}}, \bibinfo {author} {\bibfnamefont
  {T.}~\bibnamefont {Mittiga}}, \bibinfo {author} {\bibfnamefont
  {S.}~\bibnamefont {Hsieh}}, \bibinfo {author} {\bibfnamefont
  {P.}~\bibnamefont {Bhattacharyya}}, \bibinfo {author} {\bibfnamefont
  {M.}~\bibnamefont {Markham}}, \bibinfo {author} {\bibfnamefont
  {D.}~\bibnamefont {Twitchen}}, \bibinfo {author} {\bibfnamefont
  {A.}~\bibnamefont {Jarmola}}, \bibinfo {author} {\bibfnamefont
  {D.}~\bibnamefont {Budker}}, \bibinfo {author} {\bibfnamefont {C.~R.}\
  \bibnamefont {Laumann}}, \bibinfo {author} {\bibfnamefont {J.~E.}\
  \bibnamefont {Moore}},\ and\ \bibinfo {author} {\bibfnamefont {N.~Y.}\
  \bibnamefont {Yao}},\ }\bibfield  {title} {\bibinfo {title} {Emergent
  hydrodynamics in a strongly interacting dipolar spin ensemble},\ }\href
  {https://doi.org/10.1038/s41586-021-03763-1} {\bibfield  {journal} {\bibinfo
  {journal} {Nature}\ }\textbf {\bibinfo {volume} {597}},\ \bibinfo {pages}
  {45} (\bibinfo {year} {2021})}\BibitemShut {NoStop}%
\bibitem [{\citenamefont {Kolkowitz}\ \emph {et~al.}(2015)\citenamefont
  {Kolkowitz}, \citenamefont {Safira}, \citenamefont {High}, \citenamefont
  {Devlin}, \citenamefont {Choi}, \citenamefont {Unterreithmeier},
  \citenamefont {Patterson}, \citenamefont {Zibrov}, \citenamefont
  {Manucharyan}, \citenamefont {Park},\ and\ \citenamefont
  {Lukin}}]{Lukin_Johnsonnoise_2015}%
  \BibitemOpen
  \bibfield  {author} {\bibinfo {author} {\bibfnamefont {S.}~\bibnamefont
  {Kolkowitz}}, \bibinfo {author} {\bibfnamefont {A.}~\bibnamefont {Safira}},
  \bibinfo {author} {\bibfnamefont {A.~A.}\ \bibnamefont {High}}, \bibinfo
  {author} {\bibfnamefont {R.~C.}\ \bibnamefont {Devlin}}, \bibinfo {author}
  {\bibfnamefont {S.}~\bibnamefont {Choi}}, \bibinfo {author} {\bibfnamefont
  {Q.~P.}\ \bibnamefont {Unterreithmeier}}, \bibinfo {author} {\bibfnamefont
  {D.}~\bibnamefont {Patterson}}, \bibinfo {author} {\bibfnamefont {A.~S.}\
  \bibnamefont {Zibrov}}, \bibinfo {author} {\bibfnamefont {V.~E.}\
  \bibnamefont {Manucharyan}}, \bibinfo {author} {\bibfnamefont
  {H.}~\bibnamefont {Park}},\ and\ \bibinfo {author} {\bibfnamefont {M.~D.}\
  \bibnamefont {Lukin}},\ }\bibfield  {title} {\bibinfo {title} {Quantum
  electronics. probing johnson noise and ballistic transport in normal metals
  with a single-spin qubit},\ }\href {https://doi.org/10.1126/science.aaa4298}
  {\bibfield  {journal} {\bibinfo  {journal} {Science}\ }\textbf {\bibinfo
  {volume} {347}},\ \bibinfo {pages} {1129} (\bibinfo {year}
  {2015})}\BibitemShut {NoStop}%
\bibitem [{\citenamefont {Ku}\ \emph {et~al.}(2020)\citenamefont {Ku},
  \citenamefont {Zhou}, \citenamefont {Li}, \citenamefont {Shin}, \citenamefont
  {Shi}, \citenamefont {Burch}, \citenamefont {Anderson}, \citenamefont
  {Pierce}, \citenamefont {Xie}, \citenamefont {Hamo}, \citenamefont {Vool},
  \citenamefont {Zhang}, \citenamefont {Casola}, \citenamefont {Taniguchi},
  \citenamefont {Watanabe}, \citenamefont {Fogler}, \citenamefont {Kim},
  \citenamefont {Yacoby},\ and\ \citenamefont
  {Walsworth}}]{Walsworth_Diracfluid_2020}%
  \BibitemOpen
  \bibfield  {author} {\bibinfo {author} {\bibfnamefont {M.~J.~H.}\
  \bibnamefont {Ku}}, \bibinfo {author} {\bibfnamefont {T.~X.}\ \bibnamefont
  {Zhou}}, \bibinfo {author} {\bibfnamefont {Q.}~\bibnamefont {Li}}, \bibinfo
  {author} {\bibfnamefont {Y.~J.}\ \bibnamefont {Shin}}, \bibinfo {author}
  {\bibfnamefont {J.~K.}\ \bibnamefont {Shi}}, \bibinfo {author} {\bibfnamefont
  {C.}~\bibnamefont {Burch}}, \bibinfo {author} {\bibfnamefont {L.~E.}\
  \bibnamefont {Anderson}}, \bibinfo {author} {\bibfnamefont {A.~T.}\
  \bibnamefont {Pierce}}, \bibinfo {author} {\bibfnamefont {Y.}~\bibnamefont
  {Xie}}, \bibinfo {author} {\bibfnamefont {A.}~\bibnamefont {Hamo}}, \bibinfo
  {author} {\bibfnamefont {U.}~\bibnamefont {Vool}}, \bibinfo {author}
  {\bibfnamefont {H.}~\bibnamefont {Zhang}}, \bibinfo {author} {\bibfnamefont
  {F.}~\bibnamefont {Casola}}, \bibinfo {author} {\bibfnamefont
  {T.}~\bibnamefont {Taniguchi}}, \bibinfo {author} {\bibfnamefont
  {K.}~\bibnamefont {Watanabe}}, \bibinfo {author} {\bibfnamefont {M.~M.}\
  \bibnamefont {Fogler}}, \bibinfo {author} {\bibfnamefont {P.}~\bibnamefont
  {Kim}}, \bibinfo {author} {\bibfnamefont {A.}~\bibnamefont {Yacoby}},\ and\
  \bibinfo {author} {\bibfnamefont {R.~L.}\ \bibnamefont {Walsworth}},\
  }\bibfield  {title} {\bibinfo {title} {Imaging viscous flow of the dirac
  fluid in graphene},\ }\href {https://doi.org/10.1038/s41586-020-2507-2}
  {\bibfield  {journal} {\bibinfo  {journal} {Nature}\ }\textbf {\bibinfo
  {volume} {583}},\ \bibinfo {pages} {537} (\bibinfo {year}
  {2020})}\BibitemShut {NoStop}%
\bibitem [{\citenamefont {Jones}\ \emph {et~al.}(2009)\citenamefont {Jones},
  \citenamefont {Karlen}, \citenamefont {Fitzsimons}, \citenamefont {Ardavan},
  \citenamefont {Benjamin}, \citenamefont {Briggs},\ and\ \citenamefont
  {Morton}}]{Jones2009}%
  \BibitemOpen
  \bibfield  {author} {\bibinfo {author} {\bibfnamefont {J.~A.}\ \bibnamefont
  {Jones}}, \bibinfo {author} {\bibfnamefont {S.~D.}\ \bibnamefont {Karlen}},
  \bibinfo {author} {\bibfnamefont {J.}~\bibnamefont {Fitzsimons}}, \bibinfo
  {author} {\bibfnamefont {A.}~\bibnamefont {Ardavan}}, \bibinfo {author}
  {\bibfnamefont {S.~C.}\ \bibnamefont {Benjamin}}, \bibinfo {author}
  {\bibfnamefont {G.~A.~D.}\ \bibnamefont {Briggs}},\ and\ \bibinfo {author}
  {\bibfnamefont {J.~J.~L.}\ \bibnamefont {Morton}},\ }\bibfield  {title}
  {\bibinfo {title} {Magnetic field sensing beyond the standard quantum limit
  using 10-spin noon states},\ }\href {https://doi.org/10.1126/science.1170730}
  {\bibfield  {journal} {\bibinfo  {journal} {Science}\ }\textbf {\bibinfo
  {volume} {324}},\ \bibinfo {pages} {1166–1168} (\bibinfo {year}
  {2009})}\BibitemShut {NoStop}%
\bibitem [{\citenamefont {Hosten}\ \emph {et~al.}(2016)\citenamefont {Hosten},
  \citenamefont {Engelsen}, \citenamefont {Krishnakumar},\ and\ \citenamefont
  {Kasevich}}]{Mark_100times_2016}%
  \BibitemOpen
  \bibfield  {author} {\bibinfo {author} {\bibfnamefont {O.}~\bibnamefont
  {Hosten}}, \bibinfo {author} {\bibfnamefont {N.~J.}\ \bibnamefont
  {Engelsen}}, \bibinfo {author} {\bibfnamefont {R.}~\bibnamefont
  {Krishnakumar}},\ and\ \bibinfo {author} {\bibfnamefont {M.~A.}\ \bibnamefont
  {Kasevich}},\ }\bibfield  {title} {\bibinfo {title} {Measurement noise 100
  times lower than the quantum-projection limit using entangled atoms},\ }\href
  {https://doi.org/10.1038/nature16176} {\bibfield  {journal} {\bibinfo
  {journal} {Nature}\ }\textbf {\bibinfo {volume} {529}},\ \bibinfo {pages}
  {505} (\bibinfo {year} {2016})}\BibitemShut {NoStop}%
\bibitem [{\citenamefont {Xie}\ \emph {et~al.}(2021)\citenamefont {Xie},
  \citenamefont {Zhao}, \citenamefont {Kong}, \citenamefont {Ma}, \citenamefont
  {Wang}, \citenamefont {Ye}, \citenamefont {Yu}, \citenamefont {Yang},
  \citenamefont {Xu}, \citenamefont {Wang}, \citenamefont {Wang}, \citenamefont
  {Shi},\ and\ \citenamefont {Du}}]{Xie2021}%
  \BibitemOpen
  \bibfield  {author} {\bibinfo {author} {\bibfnamefont {T.}~\bibnamefont
  {Xie}}, \bibinfo {author} {\bibfnamefont {Z.}~\bibnamefont {Zhao}}, \bibinfo
  {author} {\bibfnamefont {X.}~\bibnamefont {Kong}}, \bibinfo {author}
  {\bibfnamefont {W.}~\bibnamefont {Ma}}, \bibinfo {author} {\bibfnamefont
  {M.}~\bibnamefont {Wang}}, \bibinfo {author} {\bibfnamefont {X.}~\bibnamefont
  {Ye}}, \bibinfo {author} {\bibfnamefont {P.}~\bibnamefont {Yu}}, \bibinfo
  {author} {\bibfnamefont {Z.}~\bibnamefont {Yang}}, \bibinfo {author}
  {\bibfnamefont {S.}~\bibnamefont {Xu}}, \bibinfo {author} {\bibfnamefont
  {P.}~\bibnamefont {Wang}}, \bibinfo {author} {\bibfnamefont {Y.}~\bibnamefont
  {Wang}}, \bibinfo {author} {\bibfnamefont {F.}~\bibnamefont {Shi}},\ and\
  \bibinfo {author} {\bibfnamefont {J.}~\bibnamefont {Du}},\ }\bibfield
  {title} {\bibinfo {title} {Beating the standard quantum limit under ambient
  conditions with solid-state spins},\ }\bibfield  {journal} {\bibinfo
  {journal} {Science Advances}\ }\textbf {\bibinfo {volume} {7}},\ \href
  {https://doi.org/10.1126/sciadv.abg9204} {10.1126/sciadv.abg9204} (\bibinfo
  {year} {2021})\BibitemShut {NoStop}%
\bibitem [{\citenamefont {Marciniak}\ \emph {et~al.}(2022)\citenamefont
  {Marciniak}, \citenamefont {Feldker}, \citenamefont {Pogorelov},
  \citenamefont {Kaubruegger}, \citenamefont {Vasilyev}, \citenamefont
  {Van~Bijnen}, \citenamefont {Schindler}, \citenamefont {Zoller},
  \citenamefont {Blatt},\ and\ \citenamefont {Monz}}]{Thomas_Metrology_2022}%
  \BibitemOpen
  \bibfield  {author} {\bibinfo {author} {\bibfnamefont {C.~D.}\ \bibnamefont
  {Marciniak}}, \bibinfo {author} {\bibfnamefont {T.}~\bibnamefont {Feldker}},
  \bibinfo {author} {\bibfnamefont {I.}~\bibnamefont {Pogorelov}}, \bibinfo
  {author} {\bibfnamefont {R.}~\bibnamefont {Kaubruegger}}, \bibinfo {author}
  {\bibfnamefont {D.~V.}\ \bibnamefont {Vasilyev}}, \bibinfo {author}
  {\bibfnamefont {R.}~\bibnamefont {Van~Bijnen}}, \bibinfo {author}
  {\bibfnamefont {P.}~\bibnamefont {Schindler}}, \bibinfo {author}
  {\bibfnamefont {P.}~\bibnamefont {Zoller}}, \bibinfo {author} {\bibfnamefont
  {R.}~\bibnamefont {Blatt}},\ and\ \bibinfo {author} {\bibfnamefont
  {T.}~\bibnamefont {Monz}},\ }\bibfield  {title} {\bibinfo {title} {Optimal
  metrology with programmable quantum sensors},\ }\href
  {https://doi.org/10.1038/s41586-022-04435-4} {\bibfield  {journal} {\bibinfo
  {journal} {Nature}\ }\textbf {\bibinfo {volume} {603}},\ \bibinfo {pages}
  {604} (\bibinfo {year} {2022})}\BibitemShut {NoStop}%
\bibitem [{\citenamefont {Romach}\ \emph {et~al.}(2015)\citenamefont {Romach},
  \citenamefont {Muller}, \citenamefont {Unden}, \citenamefont {Rogers},
  \citenamefont {Isoda}, \citenamefont {Itoh}, \citenamefont {Markham},
  \citenamefont {Stacey}, \citenamefont {Meijer}, \citenamefont {Pezzagna},
  \citenamefont {Naydenov}, \citenamefont {McGuinness}, \citenamefont
  {Bar-Gill},\ and\ \citenamefont {Jelezko}}]{Jelezko_surface_2014}%
  \BibitemOpen
  \bibfield  {author} {\bibinfo {author} {\bibfnamefont {Y.}~\bibnamefont
  {Romach}}, \bibinfo {author} {\bibfnamefont {C.}~\bibnamefont {Muller}},
  \bibinfo {author} {\bibfnamefont {T.}~\bibnamefont {Unden}}, \bibinfo
  {author} {\bibfnamefont {L.~J.}\ \bibnamefont {Rogers}}, \bibinfo {author}
  {\bibfnamefont {T.}~\bibnamefont {Isoda}}, \bibinfo {author} {\bibfnamefont
  {K.~M.}\ \bibnamefont {Itoh}}, \bibinfo {author} {\bibfnamefont
  {M.}~\bibnamefont {Markham}}, \bibinfo {author} {\bibfnamefont
  {A.}~\bibnamefont {Stacey}}, \bibinfo {author} {\bibfnamefont
  {J.}~\bibnamefont {Meijer}}, \bibinfo {author} {\bibfnamefont
  {S.}~\bibnamefont {Pezzagna}}, \bibinfo {author} {\bibfnamefont
  {B.}~\bibnamefont {Naydenov}}, \bibinfo {author} {\bibfnamefont {L.~P.}\
  \bibnamefont {McGuinness}}, \bibinfo {author} {\bibfnamefont
  {N.}~\bibnamefont {Bar-Gill}},\ and\ \bibinfo {author} {\bibfnamefont
  {F.}~\bibnamefont {Jelezko}},\ }\bibfield  {title} {\bibinfo {title}
  {Spectroscopy of surface-induced noise using shallow spins in diamond},\
  }\href {https://doi.org/10.1103/PhysRevLett.114.017601} {\bibfield  {journal}
  {\bibinfo  {journal} {Phys Rev Lett}\ }\textbf {\bibinfo {volume} {114}},\
  \bibinfo {pages} {017601} (\bibinfo {year} {2015})}\BibitemShut {NoStop}%
\bibitem [{\citenamefont {Rosskopf}\ \emph {et~al.}(2014)\citenamefont
  {Rosskopf}, \citenamefont {Dussaux}, \citenamefont {Ohashi}, \citenamefont
  {Loretz}, \citenamefont {Schirhagl}, \citenamefont {Watanabe}, \citenamefont
  {Shikata}, \citenamefont {Itoh},\ and\ \citenamefont
  {Degen}}]{Degen_surfacenoise_2014}%
  \BibitemOpen
  \bibfield  {author} {\bibinfo {author} {\bibfnamefont {T.}~\bibnamefont
  {Rosskopf}}, \bibinfo {author} {\bibfnamefont {A.}~\bibnamefont {Dussaux}},
  \bibinfo {author} {\bibfnamefont {K.}~\bibnamefont {Ohashi}}, \bibinfo
  {author} {\bibfnamefont {M.}~\bibnamefont {Loretz}}, \bibinfo {author}
  {\bibfnamefont {R.}~\bibnamefont {Schirhagl}}, \bibinfo {author}
  {\bibfnamefont {H.}~\bibnamefont {Watanabe}}, \bibinfo {author}
  {\bibfnamefont {S.}~\bibnamefont {Shikata}}, \bibinfo {author} {\bibfnamefont
  {K.~M.}\ \bibnamefont {Itoh}},\ and\ \bibinfo {author} {\bibfnamefont
  {C.~L.}\ \bibnamefont {Degen}},\ }\bibfield  {title} {\bibinfo {title}
  {Investigation of surface magnetic noise by shallow spins in diamond},\
  }\href {https://doi.org/10.1103/PhysRevLett.112.147602} {\bibfield  {journal}
  {\bibinfo  {journal} {Phys Rev Lett}\ }\textbf {\bibinfo {volume} {112}},\
  \bibinfo {pages} {147602} (\bibinfo {year} {2014})}\BibitemShut {NoStop}%
\bibitem [{\citenamefont {Myers}\ \emph {et~al.}(2014)\citenamefont {Myers},
  \citenamefont {Das}, \citenamefont {Dartiailh}, \citenamefont {Ohno},
  \citenamefont {Awschalom},\ and\ \citenamefont
  {Bleszynski~Jayich}}]{Jayich_surfacenoise_2014}%
  \BibitemOpen
  \bibfield  {author} {\bibinfo {author} {\bibfnamefont {B.~A.}\ \bibnamefont
  {Myers}}, \bibinfo {author} {\bibfnamefont {A.}~\bibnamefont {Das}}, \bibinfo
  {author} {\bibfnamefont {M.~C.}\ \bibnamefont {Dartiailh}}, \bibinfo {author}
  {\bibfnamefont {K.}~\bibnamefont {Ohno}}, \bibinfo {author} {\bibfnamefont
  {D.~D.}\ \bibnamefont {Awschalom}},\ and\ \bibinfo {author} {\bibfnamefont
  {A.~C.}\ \bibnamefont {Bleszynski~Jayich}},\ }\bibfield  {title} {\bibinfo
  {title} {Probing surface noise with depth-calibrated spins in diamond},\
  }\href {https://doi.org/10.1103/PhysRevLett.113.027602} {\bibfield  {journal}
  {\bibinfo  {journal} {Phys Rev Lett}\ }\textbf {\bibinfo {volume} {113}},\
  \bibinfo {pages} {027602} (\bibinfo {year} {2014})}\BibitemShut {NoStop}%
\bibitem [{\citenamefont {Barry}\ \emph {et~al.}(2020)\citenamefont {Barry},
  \citenamefont {Schloss}, \citenamefont {Bauch}, \citenamefont {Turner},
  \citenamefont {Hart}, \citenamefont {Pham},\ and\ \citenamefont
  {Walsworth}}]{Barry2020}%
  \BibitemOpen
  \bibfield  {author} {\bibinfo {author} {\bibfnamefont {J.~F.}\ \bibnamefont
  {Barry}}, \bibinfo {author} {\bibfnamefont {J.~M.}\ \bibnamefont {Schloss}},
  \bibinfo {author} {\bibfnamefont {E.}~\bibnamefont {Bauch}}, \bibinfo
  {author} {\bibfnamefont {M.~J.}\ \bibnamefont {Turner}}, \bibinfo {author}
  {\bibfnamefont {C.~A.}\ \bibnamefont {Hart}}, \bibinfo {author}
  {\bibfnamefont {L.~M.}\ \bibnamefont {Pham}},\ and\ \bibinfo {author}
  {\bibfnamefont {R.~L.}\ \bibnamefont {Walsworth}},\ }\bibfield  {title}
  {\bibinfo {title} {Sensitivity optimization for nv-diamond magnetometry},\
  }\bibfield  {journal} {\bibinfo  {journal} {Reviews of Modern Physics}\
  }\textbf {\bibinfo {volume} {92}},\ \href
  {https://doi.org/10.1103/revmodphys.92.015004} {10.1103/revmodphys.92.015004}
  (\bibinfo {year} {2020})\BibitemShut {NoStop}%
\bibitem [{\citenamefont {Cooper}\ \emph {et~al.}(2019)\citenamefont {Cooper},
  \citenamefont {Sun}, \citenamefont {Jaskula},\ and\ \citenamefont
  {Cappellaro}}]{Cooper2019}%
  \BibitemOpen
  \bibfield  {author} {\bibinfo {author} {\bibfnamefont {A.}~\bibnamefont
  {Cooper}}, \bibinfo {author} {\bibfnamefont {W.~K.~C.}\ \bibnamefont {Sun}},
  \bibinfo {author} {\bibfnamefont {J.-C.}\ \bibnamefont {Jaskula}},\ and\
  \bibinfo {author} {\bibfnamefont {P.}~\bibnamefont {Cappellaro}},\ }\bibfield
   {title} {\bibinfo {title} {Environment-assisted quantum-enhanced sensing
  with electronic spins in diamond},\ }\bibfield  {journal} {\bibinfo
  {journal} {Physical Review Applied}\ }\textbf {\bibinfo {volume} {12}},\
  \href {https://doi.org/10.1103/physrevapplied.12.044047}
  {10.1103/physrevapplied.12.044047} (\bibinfo {year} {2019})\BibitemShut
  {NoStop}%
\bibitem [{\citenamefont {Koehl}\ \emph {et~al.}(2011)\citenamefont {Koehl},
  \citenamefont {Buckley}, \citenamefont {Heremans}, \citenamefont {Calusine},\
  and\ \citenamefont {Awschalom}}]{Awschalom_SiC_2011}%
  \BibitemOpen
  \bibfield  {author} {\bibinfo {author} {\bibfnamefont {W.~F.}\ \bibnamefont
  {Koehl}}, \bibinfo {author} {\bibfnamefont {B.~B.}\ \bibnamefont {Buckley}},
  \bibinfo {author} {\bibfnamefont {F.~J.}\ \bibnamefont {Heremans}}, \bibinfo
  {author} {\bibfnamefont {G.}~\bibnamefont {Calusine}},\ and\ \bibinfo
  {author} {\bibfnamefont {D.~D.}\ \bibnamefont {Awschalom}},\ }\bibfield
  {title} {\bibinfo {title} {Room temperature coherent control of defect spin
  qubits in silicon carbide},\ }\href {https://doi.org/10.1038/nature10562}
  {\bibfield  {journal} {\bibinfo  {journal} {Nature}\ }\textbf {\bibinfo
  {volume} {479}},\ \bibinfo {pages} {84} (\bibinfo {year} {2011})}\BibitemShut
  {NoStop}%
\bibitem [{\citenamefont {Babin}\ \emph {et~al.}(2022)\citenamefont {Babin},
  \citenamefont {Stöhr}, \citenamefont {Morioka}, \citenamefont {Linkewitz},
  \citenamefont {Steidl}, \citenamefont {Wörnle}, \citenamefont {Liu},
  \citenamefont {Hesselmeier}, \citenamefont {Vorobyov}, \citenamefont
  {Denisenko}, \citenamefont {Hentschel}, \citenamefont {Gobert}, \citenamefont
  {Berwian}, \citenamefont {Astakhov}, \citenamefont {Knolle}, \citenamefont
  {Majety}, \citenamefont {Saha}, \citenamefont {Radulaski}, \citenamefont
  {Son}, \citenamefont {Ul-Hassan}, \citenamefont {Kaiser},\ and\ \citenamefont
  {Wrachtrup}}]{Joerg_SiC_2024}%
  \BibitemOpen
  \bibfield  {author} {\bibinfo {author} {\bibfnamefont {C.}~\bibnamefont
  {Babin}}, \bibinfo {author} {\bibfnamefont {R.}~\bibnamefont {Stöhr}},
  \bibinfo {author} {\bibfnamefont {N.}~\bibnamefont {Morioka}}, \bibinfo
  {author} {\bibfnamefont {T.}~\bibnamefont {Linkewitz}}, \bibinfo {author}
  {\bibfnamefont {T.}~\bibnamefont {Steidl}}, \bibinfo {author} {\bibfnamefont
  {R.}~\bibnamefont {Wörnle}}, \bibinfo {author} {\bibfnamefont
  {D.}~\bibnamefont {Liu}}, \bibinfo {author} {\bibfnamefont {E.}~\bibnamefont
  {Hesselmeier}}, \bibinfo {author} {\bibfnamefont {V.}~\bibnamefont
  {Vorobyov}}, \bibinfo {author} {\bibfnamefont {A.}~\bibnamefont {Denisenko}},
  \bibinfo {author} {\bibfnamefont {M.}~\bibnamefont {Hentschel}}, \bibinfo
  {author} {\bibfnamefont {C.}~\bibnamefont {Gobert}}, \bibinfo {author}
  {\bibfnamefont {P.}~\bibnamefont {Berwian}}, \bibinfo {author} {\bibfnamefont
  {G.~V.}\ \bibnamefont {Astakhov}}, \bibinfo {author} {\bibfnamefont
  {W.}~\bibnamefont {Knolle}}, \bibinfo {author} {\bibfnamefont
  {S.}~\bibnamefont {Majety}}, \bibinfo {author} {\bibfnamefont
  {P.}~\bibnamefont {Saha}}, \bibinfo {author} {\bibfnamefont {M.}~\bibnamefont
  {Radulaski}}, \bibinfo {author} {\bibfnamefont {N.~T.}\ \bibnamefont {Son}},
  \bibinfo {author} {\bibfnamefont {J.}~\bibnamefont {Ul-Hassan}}, \bibinfo
  {author} {\bibfnamefont {F.}~\bibnamefont {Kaiser}},\ and\ \bibinfo {author}
  {\bibfnamefont {J.}~\bibnamefont {Wrachtrup}},\ }\bibfield  {title} {\bibinfo
  {title} {Fabrication and nanophotonic waveguide integration of silicon
  carbide colour centres with preserved spin-optical coherence},\ }\href
  {https://doi.org/10.1038/s41563-021-01148-3} {\bibfield  {journal} {\bibinfo
  {journal} {Nature Materials}\ }\textbf {\bibinfo {volume} {21}},\ \bibinfo
  {pages} {67} (\bibinfo {year} {2022})}\BibitemShut {NoStop}%
\bibitem [{\citenamefont {Esat}\ \emph {et~al.}(2024)\citenamefont {Esat},
  \citenamefont {Borodin}, \citenamefont {Oh}, \citenamefont {Heinrich},
  \citenamefont {Tautz}, \citenamefont {Bae},\ and\ \citenamefont
  {Temirov}}]{Benjiamin_molecule_2024}%
  \BibitemOpen
  \bibfield  {author} {\bibinfo {author} {\bibfnamefont {T.}~\bibnamefont
  {Esat}}, \bibinfo {author} {\bibfnamefont {D.}~\bibnamefont {Borodin}},
  \bibinfo {author} {\bibfnamefont {J.}~\bibnamefont {Oh}}, \bibinfo {author}
  {\bibfnamefont {A.~J.}\ \bibnamefont {Heinrich}}, \bibinfo {author}
  {\bibfnamefont {F.~S.}\ \bibnamefont {Tautz}}, \bibinfo {author}
  {\bibfnamefont {Y.}~\bibnamefont {Bae}},\ and\ \bibinfo {author}
  {\bibfnamefont {R.}~\bibnamefont {Temirov}},\ }\bibfield  {title} {\bibinfo
  {title} {A quantum sensor for atomic-scale electric and magnetic fields},\
  }\href {https://doi.org/10.1038/s41565-024-01724-z} {\bibfield  {journal}
  {\bibinfo  {journal} {Nature Nanotechnology}\ }\textbf {\bibinfo {volume}
  {19}},\ \bibinfo {pages} {1466} (\bibinfo {year} {2024})}\BibitemShut
  {NoStop}%
\bibitem [{\citenamefont {Balasubramanian}\ \emph {et~al.}(2008)\citenamefont
  {Balasubramanian}, \citenamefont {Chan}, \citenamefont {Kolesov},
  \citenamefont {Al-Hmoud}, \citenamefont {Tisler}, \citenamefont {Shin},
  \citenamefont {Kim}, \citenamefont {Wojcik}, \citenamefont {Hemmer},
  \citenamefont {Krueger}, \citenamefont {Hanke}, \citenamefont
  {Leitenstorfer}, \citenamefont {Bratschitsch}, \citenamefont {Jelezko},\ and\
  \citenamefont {Wrachtrup}}]{Balasubramanian2008}%
  \BibitemOpen
  \bibfield  {author} {\bibinfo {author} {\bibfnamefont {G.}~\bibnamefont
  {Balasubramanian}}, \bibinfo {author} {\bibfnamefont {I.~Y.}\ \bibnamefont
  {Chan}}, \bibinfo {author} {\bibfnamefont {R.}~\bibnamefont {Kolesov}},
  \bibinfo {author} {\bibfnamefont {M.}~\bibnamefont {Al-Hmoud}}, \bibinfo
  {author} {\bibfnamefont {J.}~\bibnamefont {Tisler}}, \bibinfo {author}
  {\bibfnamefont {C.}~\bibnamefont {Shin}}, \bibinfo {author} {\bibfnamefont
  {C.}~\bibnamefont {Kim}}, \bibinfo {author} {\bibfnamefont {A.}~\bibnamefont
  {Wojcik}}, \bibinfo {author} {\bibfnamefont {P.~R.}\ \bibnamefont {Hemmer}},
  \bibinfo {author} {\bibfnamefont {A.}~\bibnamefont {Krueger}}, \bibinfo
  {author} {\bibfnamefont {T.}~\bibnamefont {Hanke}}, \bibinfo {author}
  {\bibfnamefont {A.}~\bibnamefont {Leitenstorfer}}, \bibinfo {author}
  {\bibfnamefont {R.}~\bibnamefont {Bratschitsch}}, \bibinfo {author}
  {\bibfnamefont {F.}~\bibnamefont {Jelezko}},\ and\ \bibinfo {author}
  {\bibfnamefont {J.}~\bibnamefont {Wrachtrup}},\ }\bibfield  {title} {\bibinfo
  {title} {Nanoscale imaging magnetometry with diamond spins under ambient
  conditions},\ }\href {https://doi.org/10.1038/nature07278} {\bibfield
  {journal} {\bibinfo  {journal} {Nature}\ }\textbf {\bibinfo {volume} {455}},\
  \bibinfo {pages} {648–651} (\bibinfo {year} {2008})}\BibitemShut {NoStop}%
\bibitem [{\citenamefont {Maze}\ \emph {et~al.}(2008)\citenamefont {Maze},
  \citenamefont {Stanwix}, \citenamefont {Hodges}, \citenamefont {Hong},
  \citenamefont {Taylor}, \citenamefont {Cappellaro}, \citenamefont {Jiang},
  \citenamefont {Dutt}, \citenamefont {Togan}, \citenamefont {Zibrov},
  \citenamefont {Yacoby}, \citenamefont {Walsworth},\ and\ \citenamefont
  {Lukin}}]{Maze2008}%
  \BibitemOpen
  \bibfield  {author} {\bibinfo {author} {\bibfnamefont {J.~R.}\ \bibnamefont
  {Maze}}, \bibinfo {author} {\bibfnamefont {P.~L.}\ \bibnamefont {Stanwix}},
  \bibinfo {author} {\bibfnamefont {J.~S.}\ \bibnamefont {Hodges}}, \bibinfo
  {author} {\bibfnamefont {S.}~\bibnamefont {Hong}}, \bibinfo {author}
  {\bibfnamefont {J.~M.}\ \bibnamefont {Taylor}}, \bibinfo {author}
  {\bibfnamefont {P.}~\bibnamefont {Cappellaro}}, \bibinfo {author}
  {\bibfnamefont {L.}~\bibnamefont {Jiang}}, \bibinfo {author} {\bibfnamefont
  {M.~V.~G.}\ \bibnamefont {Dutt}}, \bibinfo {author} {\bibfnamefont
  {E.}~\bibnamefont {Togan}}, \bibinfo {author} {\bibfnamefont {A.~S.}\
  \bibnamefont {Zibrov}}, \bibinfo {author} {\bibfnamefont {A.}~\bibnamefont
  {Yacoby}}, \bibinfo {author} {\bibfnamefont {R.~L.}\ \bibnamefont
  {Walsworth}},\ and\ \bibinfo {author} {\bibfnamefont {M.~D.}\ \bibnamefont
  {Lukin}},\ }\bibfield  {title} {\bibinfo {title} {Nanoscale magnetic sensing
  with an individual electronic spin in diamond},\ }\href
  {https://doi.org/10.1038/nature07279} {\bibfield  {journal} {\bibinfo
  {journal} {Nature}\ }\textbf {\bibinfo {volume} {455}},\ \bibinfo {pages}
  {644–647} (\bibinfo {year} {2008})}\BibitemShut {NoStop}%
\bibitem [{\citenamefont {Zhao}\ \emph {et~al.}(2012)\citenamefont {Zhao},
  \citenamefont {Honert}, \citenamefont {Schmid}, \citenamefont {Klas},
  \citenamefont {Isoya}, \citenamefont {Markham}, \citenamefont {Twitchen},
  \citenamefont {Jelezko}, \citenamefont {Liu}, \citenamefont {Fedder},\ and\
  \citenamefont {Wrachtrup}}]{Renbao_nspin_2012}%
  \BibitemOpen
  \bibfield  {author} {\bibinfo {author} {\bibfnamefont {N.}~\bibnamefont
  {Zhao}}, \bibinfo {author} {\bibfnamefont {J.}~\bibnamefont {Honert}},
  \bibinfo {author} {\bibfnamefont {B.}~\bibnamefont {Schmid}}, \bibinfo
  {author} {\bibfnamefont {M.}~\bibnamefont {Klas}}, \bibinfo {author}
  {\bibfnamefont {J.}~\bibnamefont {Isoya}}, \bibinfo {author} {\bibfnamefont
  {M.}~\bibnamefont {Markham}}, \bibinfo {author} {\bibfnamefont
  {D.}~\bibnamefont {Twitchen}}, \bibinfo {author} {\bibfnamefont
  {F.}~\bibnamefont {Jelezko}}, \bibinfo {author} {\bibfnamefont {R.~B.}\
  \bibnamefont {Liu}}, \bibinfo {author} {\bibfnamefont {H.}~\bibnamefont
  {Fedder}},\ and\ \bibinfo {author} {\bibfnamefont {J.}~\bibnamefont
  {Wrachtrup}},\ }\bibfield  {title} {\bibinfo {title} {Sensing single remote
  nuclear spins},\ }\href {https://doi.org/10.1038/nnano.2012.152} {\bibfield
  {journal} {\bibinfo  {journal} {Nat Nanotechnol}\ }\textbf {\bibinfo {volume}
  {7}},\ \bibinfo {pages} {657} (\bibinfo {year} {2012})}\BibitemShut {NoStop}%
\bibitem [{\citenamefont {Staudacher}\ \emph {et~al.}(2013)\citenamefont
  {Staudacher}, \citenamefont {Shi}, \citenamefont {Pezzagna}, \citenamefont
  {Meijer}, \citenamefont {Du}, \citenamefont {Meriles}, \citenamefont
  {Reinhard},\ and\ \citenamefont {Wrachtrup}}]{Staudacher2013}%
  \BibitemOpen
  \bibfield  {author} {\bibinfo {author} {\bibfnamefont {T.}~\bibnamefont
  {Staudacher}}, \bibinfo {author} {\bibfnamefont {F.}~\bibnamefont {Shi}},
  \bibinfo {author} {\bibfnamefont {S.}~\bibnamefont {Pezzagna}}, \bibinfo
  {author} {\bibfnamefont {J.}~\bibnamefont {Meijer}}, \bibinfo {author}
  {\bibfnamefont {J.}~\bibnamefont {Du}}, \bibinfo {author} {\bibfnamefont
  {C.~A.}\ \bibnamefont {Meriles}}, \bibinfo {author} {\bibfnamefont
  {F.}~\bibnamefont {Reinhard}},\ and\ \bibinfo {author} {\bibfnamefont
  {J.}~\bibnamefont {Wrachtrup}},\ }\bibfield  {title} {\bibinfo {title}
  {Nuclear magnetic resonance spectroscopy on a (5-nanometer) 3 sample
  volume},\ }\href {https://doi.org/10.1126/science.1231675} {\bibfield
  {journal} {\bibinfo  {journal} {Science}\ }\textbf {\bibinfo {volume}
  {339}},\ \bibinfo {pages} {561–563} (\bibinfo {year} {2013})}\BibitemShut
  {NoStop}%
\bibitem [{\citenamefont {Rugar}\ \emph {et~al.}(2014)\citenamefont {Rugar},
  \citenamefont {Mamin}, \citenamefont {Sherwood}, \citenamefont {Kim},
  \citenamefont {Rettner}, \citenamefont {Ohno},\ and\ \citenamefont
  {Awschalom}}]{Rugar2014}%
  \BibitemOpen
  \bibfield  {author} {\bibinfo {author} {\bibfnamefont {D.}~\bibnamefont
  {Rugar}}, \bibinfo {author} {\bibfnamefont {H.~J.}\ \bibnamefont {Mamin}},
  \bibinfo {author} {\bibfnamefont {M.~H.}\ \bibnamefont {Sherwood}}, \bibinfo
  {author} {\bibfnamefont {M.}~\bibnamefont {Kim}}, \bibinfo {author}
  {\bibfnamefont {C.~T.}\ \bibnamefont {Rettner}}, \bibinfo {author}
  {\bibfnamefont {K.}~\bibnamefont {Ohno}},\ and\ \bibinfo {author}
  {\bibfnamefont {D.~D.}\ \bibnamefont {Awschalom}},\ }\bibfield  {title}
  {\bibinfo {title} {Proton magnetic resonance imaging using a
  nitrogen–vacancy spin sensor},\ }\href
  {https://doi.org/10.1038/nnano.2014.288} {\bibfield  {journal} {\bibinfo
  {journal} {Nature Nanotechnology}\ }\textbf {\bibinfo {volume} {10}},\
  \bibinfo {pages} {120–124} (\bibinfo {year} {2014})}\BibitemShut {NoStop}%
\bibitem [{\citenamefont {Shi}\ \emph {et~al.}(2015)\citenamefont {Shi},
  \citenamefont {Zhang}, \citenamefont {Wang}, \citenamefont {Sun},
  \citenamefont {Wang}, \citenamefont {Rong}, \citenamefont {Chen},
  \citenamefont {Ju}, \citenamefont {Reinhard}, \citenamefont {Chen},
  \citenamefont {Wrachtrup}, \citenamefont {Wang},\ and\ \citenamefont
  {Du}}]{Shi2015}%
  \BibitemOpen
  \bibfield  {author} {\bibinfo {author} {\bibfnamefont {F.}~\bibnamefont
  {Shi}}, \bibinfo {author} {\bibfnamefont {Q.}~\bibnamefont {Zhang}}, \bibinfo
  {author} {\bibfnamefont {P.}~\bibnamefont {Wang}}, \bibinfo {author}
  {\bibfnamefont {H.}~\bibnamefont {Sun}}, \bibinfo {author} {\bibfnamefont
  {J.}~\bibnamefont {Wang}}, \bibinfo {author} {\bibfnamefont {X.}~\bibnamefont
  {Rong}}, \bibinfo {author} {\bibfnamefont {M.}~\bibnamefont {Chen}}, \bibinfo
  {author} {\bibfnamefont {C.}~\bibnamefont {Ju}}, \bibinfo {author}
  {\bibfnamefont {F.}~\bibnamefont {Reinhard}}, \bibinfo {author}
  {\bibfnamefont {H.}~\bibnamefont {Chen}}, \bibinfo {author} {\bibfnamefont
  {J.}~\bibnamefont {Wrachtrup}}, \bibinfo {author} {\bibfnamefont
  {J.}~\bibnamefont {Wang}},\ and\ \bibinfo {author} {\bibfnamefont
  {J.}~\bibnamefont {Du}},\ }\bibfield  {title} {\bibinfo {title}
  {Single-protein spin resonance spectroscopy under ambient conditions},\
  }\href {https://doi.org/10.1126/science.aaa2253} {\bibfield  {journal}
  {\bibinfo  {journal} {Science}\ }\textbf {\bibinfo {volume} {347}},\ \bibinfo
  {pages} {1135–1138} (\bibinfo {year} {2015})}\BibitemShut {NoStop}%
\bibitem [{\citenamefont {Sushkov}\ \emph
  {et~al.}(2014{\natexlab{a}})\citenamefont {Sushkov}, \citenamefont
  {Chisholm}, \citenamefont {Lovchinsky}, \citenamefont {Kubo}, \citenamefont
  {Lo}, \citenamefont {Bennett}, \citenamefont {Hunger}, \citenamefont
  {Akimov}, \citenamefont {Walsworth}, \citenamefont {Park},\ and\
  \citenamefont {Lukin}}]{Sushkov2014}%
  \BibitemOpen
  \bibfield  {author} {\bibinfo {author} {\bibfnamefont {A.~O.}\ \bibnamefont
  {Sushkov}}, \bibinfo {author} {\bibfnamefont {N.}~\bibnamefont {Chisholm}},
  \bibinfo {author} {\bibfnamefont {I.}~\bibnamefont {Lovchinsky}}, \bibinfo
  {author} {\bibfnamefont {M.}~\bibnamefont {Kubo}}, \bibinfo {author}
  {\bibfnamefont {P.~K.}\ \bibnamefont {Lo}}, \bibinfo {author} {\bibfnamefont
  {S.~D.}\ \bibnamefont {Bennett}}, \bibinfo {author} {\bibfnamefont
  {D.}~\bibnamefont {Hunger}}, \bibinfo {author} {\bibfnamefont
  {A.}~\bibnamefont {Akimov}}, \bibinfo {author} {\bibfnamefont {R.~L.}\
  \bibnamefont {Walsworth}}, \bibinfo {author} {\bibfnamefont {H.}~\bibnamefont
  {Park}},\ and\ \bibinfo {author} {\bibfnamefont {M.~D.}\ \bibnamefont
  {Lukin}},\ }\bibfield  {title} {\bibinfo {title} {All-optical sensing of a
  single-molecule electron spin},\ }\href {https://doi.org/10.1021/nl502988n}
  {\bibfield  {journal} {\bibinfo  {journal} {Nano Letters}\ }\textbf {\bibinfo
  {volume} {14}},\ \bibinfo {pages} {6443–6448} (\bibinfo {year}
  {2014}{\natexlab{a}})}\BibitemShut {NoStop}%
\bibitem [{\citenamefont {Kucsko}\ \emph {et~al.}(2013)\citenamefont {Kucsko},
  \citenamefont {Maurer}, \citenamefont {Yao}, \citenamefont {Kubo},
  \citenamefont {Noh}, \citenamefont {Lo}, \citenamefont {Park},\ and\
  \citenamefont {Lukin}}]{Kucsko2013}%
  \BibitemOpen
  \bibfield  {author} {\bibinfo {author} {\bibfnamefont {G.}~\bibnamefont
  {Kucsko}}, \bibinfo {author} {\bibfnamefont {P.~C.}\ \bibnamefont {Maurer}},
  \bibinfo {author} {\bibfnamefont {N.~Y.}\ \bibnamefont {Yao}}, \bibinfo
  {author} {\bibfnamefont {M.}~\bibnamefont {Kubo}}, \bibinfo {author}
  {\bibfnamefont {H.~J.}\ \bibnamefont {Noh}}, \bibinfo {author} {\bibfnamefont
  {P.~K.}\ \bibnamefont {Lo}}, \bibinfo {author} {\bibfnamefont
  {H.}~\bibnamefont {Park}},\ and\ \bibinfo {author} {\bibfnamefont {M.~D.}\
  \bibnamefont {Lukin}},\ }\bibfield  {title} {\bibinfo {title}
  {Nanometre-scale thermometry in a living cell},\ }\href
  {https://doi.org/10.1038/nature12373} {\bibfield  {journal} {\bibinfo
  {journal} {Nature}\ }\textbf {\bibinfo {volume} {500}},\ \bibinfo {pages}
  {54–58} (\bibinfo {year} {2013})}\BibitemShut {NoStop}%
\bibitem [{\citenamefont {Degen}\ \emph {et~al.}(2017)\citenamefont {Degen},
  \citenamefont {Reinhard},\ and\ \citenamefont {Cappellaro}}]{Degen2017}%
  \BibitemOpen
  \bibfield  {author} {\bibinfo {author} {\bibfnamefont {C.~L.}\ \bibnamefont
  {Degen}}, \bibinfo {author} {\bibfnamefont {F.}~\bibnamefont {Reinhard}},\
  and\ \bibinfo {author} {\bibfnamefont {P.}~\bibnamefont {Cappellaro}},\
  }\bibfield  {title} {\bibinfo {title} {Quantum sensing},\ }\bibfield
  {journal} {\bibinfo  {journal} {Reviews of Modern Physics}\ }\textbf
  {\bibinfo {volume} {89}},\ \href
  {https://doi.org/10.1103/revmodphys.89.035002} {10.1103/revmodphys.89.035002}
  (\bibinfo {year} {2017})\BibitemShut {NoStop}%
\bibitem [{\citenamefont {Wang}\ \emph {et~al.}(2024)\citenamefont {Wang},
  \citenamefont {Wang}, \citenamefont {Liu}, \citenamefont {Xu}, \citenamefont
  {Yang}, \citenamefont {Yu}, \citenamefont {Sun}, \citenamefont {Ye},
  \citenamefont {Zhou}, \citenamefont {Goncharov}, \citenamefont {Wang},\ and\
  \citenamefont {Du}}]{Ya_Highpressure_2024}%
  \BibitemOpen
  \bibfield  {author} {\bibinfo {author} {\bibfnamefont {M.}~\bibnamefont
  {Wang}}, \bibinfo {author} {\bibfnamefont {Y.}~\bibnamefont {Wang}}, \bibinfo
  {author} {\bibfnamefont {Z.}~\bibnamefont {Liu}}, \bibinfo {author}
  {\bibfnamefont {G.}~\bibnamefont {Xu}}, \bibinfo {author} {\bibfnamefont
  {B.}~\bibnamefont {Yang}}, \bibinfo {author} {\bibfnamefont {P.}~\bibnamefont
  {Yu}}, \bibinfo {author} {\bibfnamefont {H.}~\bibnamefont {Sun}}, \bibinfo
  {author} {\bibfnamefont {X.}~\bibnamefont {Ye}}, \bibinfo {author}
  {\bibfnamefont {J.}~\bibnamefont {Zhou}}, \bibinfo {author} {\bibfnamefont
  {A.~F.}\ \bibnamefont {Goncharov}}, \bibinfo {author} {\bibfnamefont
  {Y.}~\bibnamefont {Wang}},\ and\ \bibinfo {author} {\bibfnamefont
  {J.}~\bibnamefont {Du}},\ }\bibfield  {title} {\bibinfo {title} {Imaging
  magnetic transition of magnetite to megabar pressures using quantum sensors
  in diamond anvil cell},\ }\href {https://doi.org/10.1038/s41467-024-52272-y}
  {\bibfield  {journal} {\bibinfo  {journal} {Nat Commun}\ }\textbf {\bibinfo
  {volume} {15}},\ \bibinfo {pages} {8843} (\bibinfo {year}
  {2024})}\BibitemShut {NoStop}%
\bibitem [{\citenamefont {Guo}\ \emph {et~al.}(2021)\citenamefont {Guo},
  \citenamefont {Wang}, \citenamefont {Wang}, \citenamefont {Wu}, \citenamefont
  {Ye}, \citenamefont {Yu}, \citenamefont {Huang}, \citenamefont {Shi},
  \citenamefont {Wang},\ and\ \citenamefont {Du}}]{Du_AFM_2021}%
  \BibitemOpen
  \bibfield  {author} {\bibinfo {author} {\bibfnamefont {M.}~\bibnamefont
  {Guo}}, \bibinfo {author} {\bibfnamefont {M.}~\bibnamefont {Wang}}, \bibinfo
  {author} {\bibfnamefont {P.}~\bibnamefont {Wang}}, \bibinfo {author}
  {\bibfnamefont {D.}~\bibnamefont {Wu}}, \bibinfo {author} {\bibfnamefont
  {X.}~\bibnamefont {Ye}}, \bibinfo {author} {\bibfnamefont {P.}~\bibnamefont
  {Yu}}, \bibinfo {author} {\bibfnamefont {Y.}~\bibnamefont {Huang}}, \bibinfo
  {author} {\bibfnamefont {F.}~\bibnamefont {Shi}}, \bibinfo {author}
  {\bibfnamefont {Y.}~\bibnamefont {Wang}},\ and\ \bibinfo {author}
  {\bibfnamefont {J.}~\bibnamefont {Du}},\ }\bibfield  {title} {\bibinfo
  {title} {A flexible nitrogen-vacancy center probe for scanning
  magnetometry},\ }\href {https://doi.org/10.1063/5.0040679} {\bibfield
  {journal} {\bibinfo  {journal} {Review of Scientific Instruments}\ }\textbf
  {\bibinfo {volume} {92}},\ \bibinfo {pages} {055001} (\bibinfo {year}
  {2021})}\BibitemShut {NoStop}%
\bibitem [{\citenamefont {Han}\ \emph {et~al.}(2025)\citenamefont {Han},
  \citenamefont {Ye}, \citenamefont {Zhou}, \citenamefont {Liu}, \citenamefont
  {Guo}, \citenamefont {Wang}, \citenamefont {Ji}, \citenamefont {Wang},\ and\
  \citenamefont {Du}}]{Han2025}%
  \BibitemOpen
  \bibfield  {author} {\bibinfo {author} {\bibfnamefont {S.}~\bibnamefont
  {Han}}, \bibinfo {author} {\bibfnamefont {X.}~\bibnamefont {Ye}}, \bibinfo
  {author} {\bibfnamefont {X.}~\bibnamefont {Zhou}}, \bibinfo {author}
  {\bibfnamefont {Z.}~\bibnamefont {Liu}}, \bibinfo {author} {\bibfnamefont
  {Y.}~\bibnamefont {Guo}}, \bibinfo {author} {\bibfnamefont {M.}~\bibnamefont
  {Wang}}, \bibinfo {author} {\bibfnamefont {W.}~\bibnamefont {Ji}}, \bibinfo
  {author} {\bibfnamefont {Y.}~\bibnamefont {Wang}},\ and\ \bibinfo {author}
  {\bibfnamefont {J.}~\bibnamefont {Du}},\ }\bibfield  {title} {\bibinfo
  {title} {Solid-state spin coherence time approaching the physical limit},\
  }\bibfield  {journal} {\bibinfo  {journal} {Science Advances}\ }\textbf
  {\bibinfo {volume} {11}},\ \href {https://doi.org/10.1126/sciadv.adr9298}
  {10.1126/sciadv.adr9298} (\bibinfo {year} {2025})\BibitemShut {NoStop}%
\bibitem [{\citenamefont {Wang}\ \emph {et~al.}(2022)\citenamefont {Wang},
  \citenamefont {Sun}, \citenamefont {Ye}, \citenamefont {Yu}, \citenamefont
  {Liu}, \citenamefont {Zhou}, \citenamefont {Wang}, \citenamefont {Shi},
  \citenamefont {Wang},\ and\ \citenamefont {Du}}]{Wang2022}%
  \BibitemOpen
  \bibfield  {author} {\bibinfo {author} {\bibfnamefont {M.}~\bibnamefont
  {Wang}}, \bibinfo {author} {\bibfnamefont {H.}~\bibnamefont {Sun}}, \bibinfo
  {author} {\bibfnamefont {X.}~\bibnamefont {Ye}}, \bibinfo {author}
  {\bibfnamefont {P.}~\bibnamefont {Yu}}, \bibinfo {author} {\bibfnamefont
  {H.}~\bibnamefont {Liu}}, \bibinfo {author} {\bibfnamefont {J.}~\bibnamefont
  {Zhou}}, \bibinfo {author} {\bibfnamefont {P.}~\bibnamefont {Wang}}, \bibinfo
  {author} {\bibfnamefont {F.}~\bibnamefont {Shi}}, \bibinfo {author}
  {\bibfnamefont {Y.}~\bibnamefont {Wang}},\ and\ \bibinfo {author}
  {\bibfnamefont {J.}~\bibnamefont {Du}},\ }\bibfield  {title} {\bibinfo
  {title} {Self-aligned patterning technique for fabricating high-performance
  diamond sensor arrays with nanoscale precision},\ }\bibfield  {journal}
  {\bibinfo  {journal} {Science Advances}\ }\textbf {\bibinfo {volume} {8}},\
  \href {https://doi.org/10.1126/sciadv.abn9573} {10.1126/sciadv.abn9573}
  (\bibinfo {year} {2022})\BibitemShut {NoStop}%
\bibitem [{\citenamefont {Dolde}\ \emph {et~al.}(2013)\citenamefont {Dolde},
  \citenamefont {Jakobi}, \citenamefont {Naydenov}, \citenamefont {Zhao},
  \citenamefont {Pezzagna}, \citenamefont {Trautmann}, \citenamefont {Meijer},
  \citenamefont {Neumann}, \citenamefont {Jelezko},\ and\ \citenamefont
  {Wrachtrup}}]{Joerg_Entanglement_2013}%
  \BibitemOpen
  \bibfield  {author} {\bibinfo {author} {\bibfnamefont {F.}~\bibnamefont
  {Dolde}}, \bibinfo {author} {\bibfnamefont {I.}~\bibnamefont {Jakobi}},
  \bibinfo {author} {\bibfnamefont {B.}~\bibnamefont {Naydenov}}, \bibinfo
  {author} {\bibfnamefont {N.}~\bibnamefont {Zhao}}, \bibinfo {author}
  {\bibfnamefont {S.}~\bibnamefont {Pezzagna}}, \bibinfo {author}
  {\bibfnamefont {C.}~\bibnamefont {Trautmann}}, \bibinfo {author}
  {\bibfnamefont {J.}~\bibnamefont {Meijer}}, \bibinfo {author} {\bibfnamefont
  {P.}~\bibnamefont {Neumann}}, \bibinfo {author} {\bibfnamefont
  {F.}~\bibnamefont {Jelezko}},\ and\ \bibinfo {author} {\bibfnamefont
  {J.}~\bibnamefont {Wrachtrup}},\ }\bibfield  {title} {\bibinfo {title}
  {Room-temperature entanglement between single defect spins in diamond},\
  }\href {https://doi.org/10.1038/nphys2545} {\bibfield  {journal} {\bibinfo
  {journal} {Nature Physics}\ }\textbf {\bibinfo {volume} {9}},\ \bibinfo
  {pages} {139} (\bibinfo {year} {2013})}\BibitemShut {NoStop}%
\bibitem [{\citenamefont {Sushkov}\ \emph
  {et~al.}(2014{\natexlab{b}})\citenamefont {Sushkov}, \citenamefont
  {Lovchinsky}, \citenamefont {Chisholm}, \citenamefont {Walsworth},
  \citenamefont {Park},\ and\ \citenamefont {Lukin}}]{Lukin_reporter_2014}%
  \BibitemOpen
  \bibfield  {author} {\bibinfo {author} {\bibfnamefont {A.~O.}\ \bibnamefont
  {Sushkov}}, \bibinfo {author} {\bibfnamefont {I.}~\bibnamefont {Lovchinsky}},
  \bibinfo {author} {\bibfnamefont {N.}~\bibnamefont {Chisholm}}, \bibinfo
  {author} {\bibfnamefont {R.~L.}\ \bibnamefont {Walsworth}}, \bibinfo {author}
  {\bibfnamefont {H.}~\bibnamefont {Park}},\ and\ \bibinfo {author}
  {\bibfnamefont {M.~D.}\ \bibnamefont {Lukin}},\ }\bibfield  {title} {\bibinfo
  {title} {Magnetic resonance detection of individual proton spins using
  quantum reporters},\ }\href {https://doi.org/10.1103/PhysRevLett.113.197601}
  {\bibfield  {journal} {\bibinfo  {journal} {Phys. Rev. Lett.}\ }\textbf
  {\bibinfo {volume} {113}},\ \bibinfo {pages} {197601} (\bibinfo {year}
  {2014}{\natexlab{b}})}\BibitemShut {NoStop}%
\bibitem [{\citenamefont {Ungar}\ \emph {et~al.}(2024)\citenamefont {Ungar},
  \citenamefont {Cappellaro}, \citenamefont {Cooper},\ and\ \citenamefont
  {Sun}}]{Ungar2024}%
  \BibitemOpen
  \bibfield  {author} {\bibinfo {author} {\bibfnamefont {A.}~\bibnamefont
  {Ungar}}, \bibinfo {author} {\bibfnamefont {P.}~\bibnamefont {Cappellaro}},
  \bibinfo {author} {\bibfnamefont {A.}~\bibnamefont {Cooper}},\ and\ \bibinfo
  {author} {\bibfnamefont {W.~K.~C.}\ \bibnamefont {Sun}},\ }\bibfield  {title}
  {\bibinfo {title} {Control of an environmental spin defect beyond the
  coherence limit of a central spin},\ }\bibfield  {journal} {\bibinfo
  {journal} {PRX Quantum}\ }\textbf {\bibinfo {volume} {5}},\ \href
  {https://doi.org/10.1103/prxquantum.5.010321} {10.1103/prxquantum.5.010321}
  (\bibinfo {year} {2024})\BibitemShut {NoStop}%
\bibitem [{\citenamefont {Stacey}\ \emph {et~al.}(2019)\citenamefont {Stacey},
  \citenamefont {Dontschuk}, \citenamefont {Chou}, \citenamefont {Broadway},
  \citenamefont {Schenk}, \citenamefont {Sear}, \citenamefont {Tetienne},
  \citenamefont {Hoffman}, \citenamefont {Prawer}, \citenamefont {Pakes},
  \citenamefont {Tadich}, \citenamefont {De~Leon}, \citenamefont {Gali},\ and\
  \citenamefont {Hollenberg}}]{Hollenberg_Surface_2019}%
  \BibitemOpen
  \bibfield  {author} {\bibinfo {author} {\bibfnamefont {A.}~\bibnamefont
  {Stacey}}, \bibinfo {author} {\bibfnamefont {N.}~\bibnamefont {Dontschuk}},
  \bibinfo {author} {\bibfnamefont {J.}~\bibnamefont {Chou}}, \bibinfo {author}
  {\bibfnamefont {D.~A.}\ \bibnamefont {Broadway}}, \bibinfo {author}
  {\bibfnamefont {A.~K.}\ \bibnamefont {Schenk}}, \bibinfo {author}
  {\bibfnamefont {M.~J.}\ \bibnamefont {Sear}}, \bibinfo {author}
  {\bibfnamefont {J.}~\bibnamefont {Tetienne}}, \bibinfo {author}
  {\bibfnamefont {A.}~\bibnamefont {Hoffman}}, \bibinfo {author} {\bibfnamefont
  {S.}~\bibnamefont {Prawer}}, \bibinfo {author} {\bibfnamefont {C.~I.}\
  \bibnamefont {Pakes}}, \bibinfo {author} {\bibfnamefont {A.}~\bibnamefont
  {Tadich}}, \bibinfo {author} {\bibfnamefont {N.~P.}\ \bibnamefont {De~Leon}},
  \bibinfo {author} {\bibfnamefont {A.}~\bibnamefont {Gali}},\ and\ \bibinfo
  {author} {\bibfnamefont {L.~C.~L.}\ \bibnamefont {Hollenberg}},\ }\bibfield
  {title} {\bibinfo {title} {Evidence for primal sp<sup>2</sup> defects at the
  diamond surface: Candidates for electron trapping and noise sources},\ }\href
  {https://doi.org/10.1002/admi.201801449} {\bibfield  {journal} {\bibinfo
  {journal} {Advanced Materials Interfaces}\ }\textbf {\bibinfo {volume} {6}},\
  \bibinfo {pages} {1801449} (\bibinfo {year} {2019})}\BibitemShut {NoStop}%
\bibitem [{\citenamefont {Murai}(2003)}]{Murai2003}%
  \BibitemOpen
  \bibfield  {author} {\bibinfo {author} {\bibfnamefont {H.}~\bibnamefont
  {Murai}},\ }\bibfield  {title} {\bibinfo {title} {Spin-chemical approach to
  photochemistry: reaction control by spin quantum operation},\ }\href
  {https://doi.org/10.1016/s1389-5567(02)00038-2} {\bibfield  {journal}
  {\bibinfo  {journal} {Journal of Photochemistry and Photobiology C:
  Photochemistry Reviews}\ }\textbf {\bibinfo {volume} {3}},\ \bibinfo {pages}
  {183–201} (\bibinfo {year} {2003})}\BibitemShut {NoStop}%
\bibitem [{\citenamefont {Huang}\ \emph {et~al.}(2025)\citenamefont {Huang},
  \citenamefont {Zhao}, \citenamefont {Kong}, \citenamefont {Wang},
  \citenamefont {Zhao}, \citenamefont {Gong}, \citenamefont {Ye}, \citenamefont
  {Wang}, \citenamefont {Shi},\ and\ \citenamefont {Du}}]{Huang2025}%
  \BibitemOpen
  \bibfield  {author} {\bibinfo {author} {\bibfnamefont {Z.}~\bibnamefont
  {Huang}}, \bibinfo {author} {\bibfnamefont {Z.}~\bibnamefont {Zhao}},
  \bibinfo {author} {\bibfnamefont {F.}~\bibnamefont {Kong}}, \bibinfo {author}
  {\bibfnamefont {Z.}~\bibnamefont {Wang}}, \bibinfo {author} {\bibfnamefont
  {P.}~\bibnamefont {Zhao}}, \bibinfo {author} {\bibfnamefont {X.}~\bibnamefont
  {Gong}}, \bibinfo {author} {\bibfnamefont {X.}~\bibnamefont {Ye}}, \bibinfo
  {author} {\bibfnamefont {Y.}~\bibnamefont {Wang}}, \bibinfo {author}
  {\bibfnamefont {F.}~\bibnamefont {Shi}},\ and\ \bibinfo {author}
  {\bibfnamefont {J.}~\bibnamefont {Du}},\ }\bibfield  {title} {\bibinfo
  {title} {Parallel accelerated electron paramagnetic resonance spectroscopy
  using diamond sensors},\ }\bibfield  {journal} {\bibinfo  {journal} {Physical
  Review Letters}\ }\textbf {\bibinfo {volume} {134}},\ \href
  {https://doi.org/10.1103/physrevlett.134.130801}
  {10.1103/physrevlett.134.130801} (\bibinfo {year} {2025})\BibitemShut
  {NoStop}%
\bibitem [{\citenamefont {Bluvstein}\ \emph {et~al.}(2019)\citenamefont
  {Bluvstein}, \citenamefont {Zhang},\ and\ \citenamefont
  {Jayich}}]{Jayich_charge_2019}%
  \BibitemOpen
  \bibfield  {author} {\bibinfo {author} {\bibfnamefont {D.}~\bibnamefont
  {Bluvstein}}, \bibinfo {author} {\bibfnamefont {Z.~R.}\ \bibnamefont
  {Zhang}},\ and\ \bibinfo {author} {\bibfnamefont {A.~C.~B.}\ \bibnamefont
  {Jayich}},\ }\bibfield  {title} {\bibinfo {title} {Identifying and mitigating
  charge instabilities in shallow diamond nitrogen-vacancy centers},\
  }\bibfield  {journal} {\bibinfo  {journal} {Physical Review Letters}\
  }\textbf {\bibinfo {volume} {122}},\ \href
  {https://doi.org/10.1103/PhysRevLett.122.076101}
  {10.1103/PhysRevLett.122.076101} (\bibinfo {year} {2019})\BibitemShut
  {NoStop}%
\end{thebibliography}%
\end{document}